\documentclass[12pt,amssymb]{article} 

\usepackage{amssymb}
\usepackage[dvips]{graphicx}

\newcommand{\newsection}[1]{
\addtocounter{section}{1} \setcounter{equation}{0}
\setcounter{subsection}{0} \addcontentsline{toc}{section}{\protect
\numberline{\arabic{section}}{{\rm #1}}} \vglue .6cm \pagebreak[3]
\noindent{\bf  \thesection. #1}\nopagebreak[4]\par\vskip .3cm}
\newcommand{\newsubsection}[1]{
\addtocounter{subsection}{1}
\addcontentsline{toc}{subsection}{\protect
\numberline{\arabic{section}.\arabic{subsection}}{#1}} \vglue .4cm
\pagebreak[3] \noindent{\it \thesubsection.
#1}\nopagebreak[4]\par\vskip .3cm}

\renewcommand{\theequation}{\thesection.\arabic{equation}}

\newlength{\extraspace}
\newlength{\extraspaces}
\newcounter{dummy}
\newcommand{\bc}{\begin{center}}
\newcommand{\ec}{\end{center}}
\newcommand{\be}{\begin{equation}
\addtolength{\abovedisplayskip}{\extraspaces}
\addtolength{\belowdisplayskip}{\extraspaces}
\addtolength{\abovedisplayshortskip}{\extraspace}
\addtolength{\belowdisplayshortskip}{\extraspace}}
\newcommand{\ee}{\end{equation}}

\newcommand{\ba}{\begin{eqnarray}
\addtolength{\abovedisplayskip}{\extraspaces}
\addtolength{\belowdisplayskip}{\extraspaces}
\addtolength{\abovedisplayshortskip}{\extraspace}
\addtolength{\belowdisplayshortskip}{\extraspace}}
\newcommand{\ea}{\end{eqnarray}}

\newcommand{\is}{& \!\! = \!\! &}
\newcommand{\ban}{\begin{eqnarray*}
\addtolength{\abovedisplayskip}{\extraspaces}
\addtolength{\belowdisplayskip}{\extraspaces}
\addtolength{\abovedisplayshortskip}{\extraspace}
\addtolength{\belowdisplayshortskip}{\extraspace}}
\newcommand{\ean}{\end{eqnarray*}}
\newcommand{\baa}{
\addtocounter{equation}{1} \setcounter{dummy}{\value{equation}}
\setcounter{equation}{0}
\renewcommand{\theequation}{\thesection.\arabic{dummy}\alph{equation}}
\begin{eqnarray}
\addtolength{\abovedisplayskip}{\extraspaces}
\addtolength{\belowdisplayskip}{\extraspaces}
\addtolength{\abovedisplayshortskip}{\extraspace}
\addtolength{\belowdisplayshortskip}{\extraspace}}
\newcommand{\eaa}{
\end{eqnarray}
\setcounter{equation}{\value{dummy}}
\renewcommand{\theequation}{\thesection.\arabic{equation}}}

\newcommand{\half}{\frac{1}{2}}
\newcommand{\del}{\partial}
\newcommand{\eol}{\nonumber \\}

\newcommand{\ch}[1]{{\rm ch}( #1 )}

\begin{document}

\begin{flushright}
March 2007\\
{\tt hep-th/0703047}\\
AEI-2007-009
\end{flushright}
\vspace{2cm}

\thispagestyle{empty}

\begin{center}
{\Large\bf  Geometry of Particle Physics
 \\[13mm] }

{\sc Martijn Wijnholt}\\[2.5mm]
{\it Max Planck Institute (Albert Einstein Institute)\\
Am M\"uhlenberg 1 \\
D-14476 Potsdam-Golm, Germany }\\
[30mm]

{\sc Abstract}

\end{center}

\noindent We explain how to construct a large class of new quiver
gauge theories from branes at singularities by orientifolding and
Higgsing old examples. The new models include the MSSM, decoupled
from gravity, as well as some classic models of dynamical SUSY
breaking. We also discuss topological criteria for unification.

  \vfill

\newpage

\renewcommand{\Large}{\normalsize}

\tableofcontents

\newpage

\newsection{Introduction}

\newsubsection{Overview: merits of local constructions}

String theory grew out of a desire to provide a framework for
particle  physics beyond the Standard Model and all the way up to
the Planck scale. In order to make progress, one needs to find an
embedding of the SM, or some realistic extension such as the MSSM,
in ten-dimensional string theory. A beautiful aspect of such a
picture is that the details of the matter content and the
interactions are governed by the geometry of field configurations in
the six additional dimensions.

The main approaches that have been considered are
\begin{itemize}
  \item Heterotic strings;
  \item Global D-brane constructions;
  \item Local D-brane constructions.
\end{itemize}
Here we have distinguished two kinds of D-brane constructions. By a
local construction we mean a construction which satisfies a {\it
correspondence principle}: we require that there is a decoupling
limit in which the 4D Planck scale goes to infinity, but the SM
couplings at some fixed energy scale remain finite. This requirement
is motivated by the existence of a large hierarchy between the TeV
scale and the Planck scale. The natural set-up which satisfies this
principle is fractional branes at a singularity.

 \begin{figure}[th]
\begin{center}
            \scalebox{.4}{
               \includegraphics[width=\textwidth]{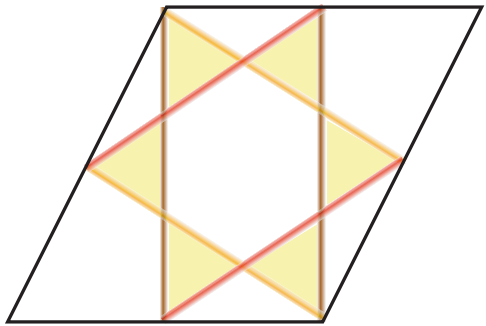}
               }
\end{center}
\vspace{-.5cm} \caption{ \it Caricature of a global D-brane model.
If the size of the $T^2$ goes to infinity, as typically happens in
the $M_{\rm pl,4} \to \infty$ limit, the volumes of the branes and
the distance between their intersections goes to infinity as well,
shutting of the Standard Model couplings.}\label{GlobalSMCartoon}
 \end{figure}
Our use of the words `local construction' differs from some of the
literature. In global constructions the SM fields are often also
localized in ten dimensions, but in the $M_{\rm Pl,4}\to\infty$
limit most of the Standard Model interactions are turned off. This
is because either the cycle on which a brane is wrapped becomes
large, turning off the gauge coupling, or because fermion and scalar
wave functions are supported on regions which get infinitely
separated in this limit, turning off Yukawa couplings. Similarly in
the heterotic string, the perturbative gauge interactions are shut
off if we take the volume of the Calabi-Yau to infinity. We will
require that all these interactions remain finite in the decoupling
limit.

We cannot guarantee that the correspondence principle is satisfied
in nature. However we believe that insisting on it is an important
model building ingredient, if only to disentangle field theoretic
model building issues from quantum gravity. In addition, insisting
on such a scenario has a number of practical advantages:
\begin{itemize}
  \item  { \it Holography}: Higher energy scales in
the gauge theory correspond to probing distances farther away from
the brane. This property allows one to take a bottom-up perspective
to model building \cite{Aldazabal:2000sa}. In order to reproduce the
SM we only need to know a local neighbourhood of the brane of radius
$r$, where $U = r/\alpha' \sim 1\, {\rm TeV}$.
 \item {\it Adjustability}: The couplings of the gauge theory
translate to boundary values of closed string fields on the boundary
of this local neighbourhood, and we may adjust them at will. Their
values are set by some high energy physics which we have not yet
included.
  \item {\it Uniqueness}: It is expected that the closed string theory
  can be recovered from the open string theory. So up to some natural
  ambiguities like T-dualities, the local neighbourhood should be completely
   determined by the ensemble of gauge theories obtained by varying the
ranks of the gauge groups. Thus finding the local geometry for a
gauge theory is a relatively well-posed problem which should have a
unique solution. The apparent non-uniqueness seen in other
approaches is reflected here in the fact that there might be many
different extensions of the same local geometry.
\end{itemize}

In \cite{Verlinde:2005jr} Herman Verlinde and the author gave a
construction of a local model resembling the MSSM.\footnote{A
closely related model was considered in \cite{Berenstein:2001nk}.}
This construction had some drawbacks which could be traced back to
the fact that we were working with oriented quivers. In this paper
we address the problem of giving a local construction of the MSSM
itself.

We have frequently seen the sentiment expressed that gauge theories
obtained from branes at singularities are somehow rather special.
The main message of this paper is not so much that we can construct
some specific models. Rather it is that with the present set of
ideas we can get pretty much any quiver gauge theory from branes at
singularities. To illustrate this point, we also engineer some classic
models of dynamical SUSY breaking.

While we touch on some more abstract topics like exceptional
collections, the strategy is really very simple. We look for an
embedding of the MSSM into a quiver gauge theory for which the
geometric description is known, and then turn on various VEVs and
mass terms. In order to keep maximum control we require the
deformations to preserve supersymmetry. Below the scale of the
masses, we can effectively integrate out and forget the extra
massive modes. On the geometric side, this corresponds to turning on
certain moduli of the fractional brane or changing the complex
structure of the singularity, and cutting off the geometry below the
scale of the superfluous massive modes. Hence we can speak of the
geometry of the MSSM.\footnote{Recently some attempts have been made
to construct such a geometry directly from the MSSM
\cite{Gray:2005sr}.}

 \begin{figure}[th]
\begin{center}
            \scalebox{.4}{
               \includegraphics[width=\textwidth]{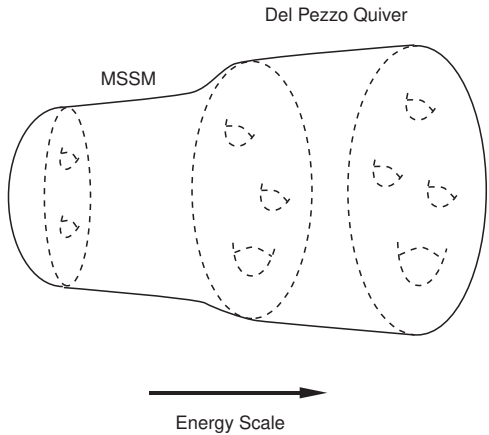}
               }
\end{center}
\vspace{-.5cm} \caption{ \it The radial direction away from the
fractional brane is interpreted as an energy scale. After Higgsing
the Del Pezzo quiver theory, a sufficiently small neighbourhood of
the singularity describes the MSSM.}\label{MSSMtoDP}
 \end{figure}
The Del Pezzo quiver and other intermediate quivers are purely
auxiliary theories which are possible UV extensions of the MSSM. Our
construction appears to be highly non-unique. This is a reflection
of the bottom-up perspective, in which the theory can be extended in
many ways beyond the TeV scale.

While we don't believe it is an issue, we should mention a possible
caveat in our construction. As we will review we can vary
superpotential terms in the original Del Pezzo quiver
independently\footnote{This is a crucial difference with generic
global D-brane models.}, and it is expected but not completely
obvious that the same is true in the Higgsed superpotential. One
would like to prove that one can vary mass terms independently so
that we can keep some non-chiral Higgs fields light and the
remainder arbitrarily heavy. We checked on the computer in a number
of simple examples that it works as expected. However in our
realistic examples some of the mass terms in the Higgsed quiver
should be induced from superpotential terms in the original Del
Pezzo quiver which are of 12th order in the fields. Unfortunately
due to memory constraints we have only been able to handle 4th and
8th order terms on the computer, and so we have not explicitly shown
in these examples that all excess non-chiral matter can be given a
mass.

\newsubsection{The MSSM as a quiver}

Let us now describe what we mean by obtaining the MSSM. With
D-branes, the best one can do is obtaining the MSSM together with an
additional massive gauge boson. In addition, the right-handed
neutrino sector is not set in stone. We first describe the quiver we
would like to produce. In later sections, we describe how to
engineer it.

Any weakly coupled\footnote{This conclusion can be evaded by using
mutually non-local 7-branes in the construction, that is by dropping
the requirement that the dilaton is small near the 7-branes.}
D-brane construction of the MSSM will have at least one extra
massive gauge boson, namely gauged baryon number\footnote{It is
possible to construct weakly coupled D-brane models in which the
extra $U(1)$ is not baryon number, eg. by taking right-handed quarks
to be in the 2-index anti-symmetric representation of $SU(3)$.
However such models are problematic at the level of interactions and
so will not be considered.}, because the $SU(3)_{\rm colour}$ always
gets enhanced to $U(3)$. In addition, we have to choose how to
realize the right-handed neutrino sector. The most likely sources
for right-handed neutrinos are:
\begin{itemize}
  \item[(a)] open strings charged with respect to a gauge symmetry that
  is not part of the SM;
  \item[(b)] uncharged open strings;
  \item[(c)] superpartners of closed string moduli.
\end{itemize}
Since all such modes are singlets under the observed low energy gauge groups,
they will probably mix and there may not be an invariant distinction between them.

\begin{figure}[th]
\begin{center}
            \scalebox{.8}{
               \includegraphics[width=\textwidth]{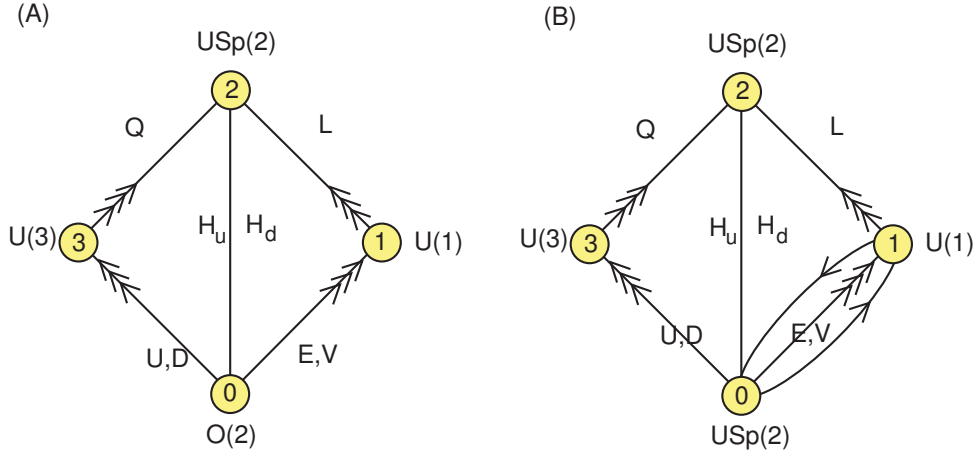}
               }
\end{center}
\vspace{-.5cm} \caption{ \it (A): An MSSM quiver, with an additional
massless $U(1)_{B-L}$. (B): Model I, a left-right unified model
which can be Higgsed to the MSSM. }\label{ECQuiver}
 \end{figure}

One of the closest quivers we could try to construct is shown in
figure \ref{ECQuiver}A.
This is essentially the `four-stack' quiver first discussed in
\cite{Ibanez:2001nd}. It consists of the MSSM plus $U(1)_B$ and
$U(1)_L$ vector bosons, and a right-handed neutrino sector from
charged open strings. Many groups have searched for this model and
closely related ones in specific compactifications, see for instance
\cite{various,Dijkstra:2004cc} and the review
\cite{Blumenhagen:2005mu}.

The combination $U(1)_{B+L}$ is anomalous, and as usual gets a mass
by coupling to a closed string axion (the St\"uckelberg mechanism).
Note this is {\it not} the PQ axion, which may or may not exist,
depending on the UV extension of the local geometry. The combination
$U(1)_{B-L}$ is not anomalous, but could still get a mass by
coupling to a closed string axion, also depending on the UV
completion. However if we take the gauge group on the bottom node to
be literally $O(2)$, i.e. obtained from an orientifold projection of
$U(2)$, then this $O(2)$ can not have a St\"uckelberg coupling to an
axion. Since we would like to keep a massless $U(1)_Y$, and since
$U(1)_Y$ is a linear combination of the $SO(2)$ and $U(1)_{B-L}$,
this means that $U(1)_{B-L}$ cannot get a mass through the
St\"uckelberg mechanism.\footnote{ This agrees with
\cite{Dijkstra:2004cc}, where all the $O(2)$ models had a massless
$U(1)_{B-L}$.}

 \begin{figure}[th]
\begin{center}
            \scalebox{.4}{
               \includegraphics[width=\textwidth]{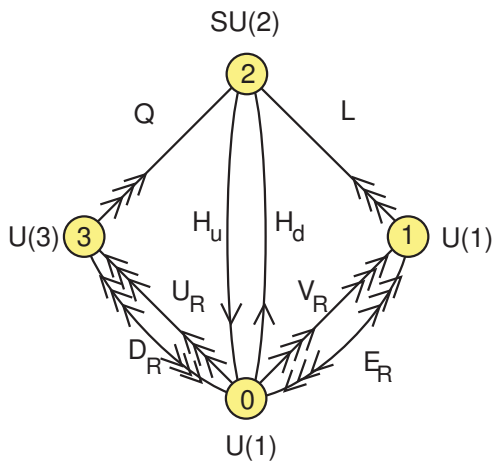}
               }
\end{center}
\vspace{-.5cm} \caption{ \it Model II: a quiver consisting of the
MSSM, $U(1)_{B-L}$ plus a massive $U(1)_{B+L}$. The $U(1)_{B-L}$ can
be coupled to a St\"uckelberg field. }\label{PITPQuiver}
 \end{figure}
Thus we have two options. Either we instead construct the
orientifold model in figure \ref{PITPQuiver}, where the $U(1)$ on
the bottom node comes from identifying two different nodes on the
covering quiver. Then we have recourse to the St\"uckelberg
mechanism to get rid of $U(1)_{B-L}$. We will call this quiver model
II. A construction of this model is given in section {\it 4.3}.

\begin{figure}[th]
\begin{center}
            \scalebox{.4}{
               \includegraphics[width=\textwidth]{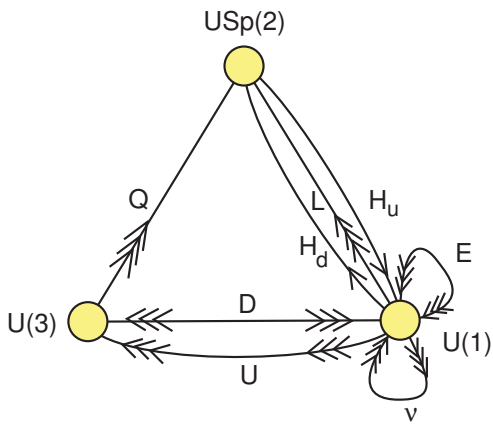}
               }
\end{center}
\vspace{-.5cm} \caption{ \it The Standard model plus $U(1)_{B}$.
Note we still need R-parity to forbid undesirable
couplings.}\label{MinQuiver}
 \end{figure}

Alternatively we can make the extra $U(1)$ massive by conventional
Higgsing. This requires adding some non-chiral matter and condensing
it, or turning on a VEV for a right handed s-neutrino. In this case,
we would finally end up with the quiver in figure
\ref{MinQuiver}.\footnote{The non-SUSY version of this quiver was
recently discussed in \cite{Berenstein:2006pk}.} This quiver
consists of the MSSM, together with a massive $U(1)_B$ gauge boson,
and a right-handed neutrino sector from uncharged open strings
(adjoints).

If in fact we use the second option, adding non-chiral matter and
Higgsing, then for our purposes here we might as well replace the
$O(2)$ with a $USp(2)$, since both break to the same model up to
some massive particles. In the local set-up, the masses of the extra
particles may be taken arbitrarily large. Moreover up to the massive
$U(1)_B$ this is actually a well known unified model, the minimal
left-right symmetric model (an intermediate step to $SO(10)$
unification), so it has some independent interest. Thus we might as
well construct the quiver in figure \ref{ECQuiver}B, which we will
call model I. This is the simpler of the constructions in this
paper, and will be explained in section {\it 4.1}.

We should point out that $R$-parity is not quite automatic in either
of our models, although both models appear to have a global
$U(1)_{B-L}$. In our first model we must preserve $R$-parity in our
final Higgsing to the MSSM. In both models we might need to worry about
$D$-instanton effects which break this symmetry after coupling to 4D
gravity, though such effects are presumably small. This is not
really surprising: the MSSM does not explain $R$-parity, it merely
assumes it. To explain it, we must know more about the UV extension
of our models.

\begin{figure}[th]
\begin{center}
            \scalebox{.8}{
               \includegraphics[width=\textwidth]{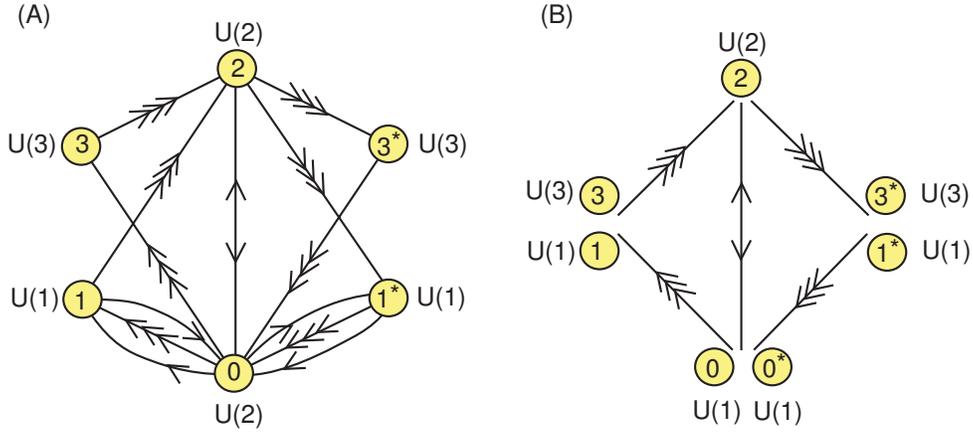}
               }
\end{center}
\vspace{-.5cm} \caption{ \it (A): Covering quiver of model I. (B):
Covering quiver of model II. In (B) we have grouped nodes together
if they have the same intersection numbers with all other nodes, to
make the diagram less cluttered.}\label{CoverQuivers}
 \end{figure}

Now to get a hint for finding these quivers, we first draw the
oriented covering quivers. The covering quiver for the quivers in
figure \ref{ECQuiver} is drawn in figure \ref{CoverQuivers}A, and
the covering quiver for figure \ref{PITPQuiver} is given in figure
\ref{CoverQuivers}B.

These quivers have to our knowledge not yet been encountered in the
literature on D-branes at singularities. Although the number of
generations doesn't match, these quivers still bear a close
resemblance to the characteristic structure of Del Pezzo quivers, and
especially to Del Pezzo 5. Recall that a Del Pezzo singularity is a Calabi-Yau
singularity with a single Del Pezzo surface collapsing to zero size.
The Del Pezzo surfaces are ${\bf P}^1 \times {\bf P}^1$ or ${\bf
P}^2$ blown up at up to eight points. We will give a discussion of
orientifolds of Del Pezzo 5 (i.e. five blow-ups of ${\bf P}^2$) in
section {\it 3}.

\newsubsection{Why Del Pezzo surfaces?}

The fact that the Del Pezzo quivers are seen to be relevant is not
surprising. It is practically guaranteed when we ask for chiral
gauge theories which are not too complicated. Let us explain this
point.

Any $CY_3$ singularity has a collection of 2- and 4-cycles
collapsing to zero size. Now chiral matter comes from intersection
of 2-cycles with 4-cycles. This is easy to see: for instance if we
have only branes wrapping 2-cycles, we can always deform the branes
(at some cost in energy) so that they don't intersect. Then all open
string modes are massive, and thus the net number of chiral fermions
must be zero. So requiring chiral fermions implies that we have to
have some collapsing 4-cycles in the geometry. The Del Pezzo
singularities, which have precisely a single collapsing 4-cycle, are
then the simplest examples.

Moreover, a minimal D-brane realization of the SM has one local
$U(1)$ which is anomalous, namely $U(1)_B$, and this lifts to two
anomalous $U(1)$'s on the oriented covering quiver. Now the the
number of anomalous $U(1)$'s is interpreted geometrically as the
rank of the intersection
matrix of vanishing cycles. Hence the Del Pezzo quivers and their
orientifolds are natural candidates because they are chiral quivers
with the minimum number of anomalous $U(1)$'s, namely two.

Although the models we are looking for are not among the known Del
Pezzo quivers, these arguments convinced us that we should derive
them from the quivers that were already known, rather than look for
new singularities.

\newpage

\newsection{Lightning review of branes at singularities}

Consider a Calabi-Yau singularity in IIb string theory, characterized by a
collection of vanishing 2- and 4-cycles. Since the curvature is very
large, it is in general not clear how to define the notion of a
D-brane at a singularity. An notable exception is the case of
orbifold singularities, where we can use free field theory. From
this special case the following picture has emerged: given a
singularity we expect the existence of a finite set of irreducible
``fractional'' branes. For the case of orbifolds these irreducible
branes are in one-to-one correspondence with the irreducible
representation of the orbifold group. To these irreducible branes we
can associate the basic quiver diagram. For each irreducible
fractional brane we draw a node, and for each massless open string
which goes from brane $i$ to brane $j$ we draw an arrow or an edge
between the corresponding nodes. All the remaining branes can be
expressed as bound states of these irreducible branes, or
equivalently as a Higgsing of the basic quiver.

\begin{figure}[th]
\begin{center}
            \scalebox{.5}{
               \includegraphics[width=\textwidth]{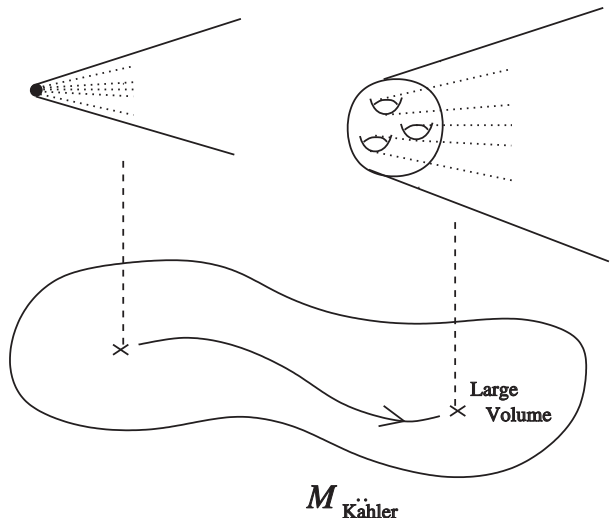}
               }
\end{center}
\vspace{-.5cm} \caption{ \it Ignoring the D-terms and extrapolating
to the large volume limit.}\label{largevolume}
 \end{figure}

Now how do we find the basic quiver for a general singularity? Let
us assume our branes are half BPS and space-time filling, so that we
get a 4D $N=1$ quiver gauge theory. Then we can use the following
strategy: we make sure that the F-term equations are satisfied, but
we temporarily ignore the D-term equations. Then we can blow up the
vanishing 2- and 4-cycles and extrapolate to the large volume limit
(figure \ref{largevolume}). This limit is unphysical from the point
of view of the quiver gauge theory, because the D-terms are not
satisfied, but in this limit we understand how to compute the F-term
equations. Moreover due to the shift symmetry of the $B$-field we
can argue that the perturbative superpotential does not depend on
complexified K\"ahler moduli and must be the same as in the small
radius limit. When the cycles are large and the curvature is small,
we can represent the D-branes by sheaves localized on the vanishing
cycles. The irreducible fractional branes get mapped to an
exceptional collection $\{ F_1, \ldots, F_n\}$, that is a collection
of rigid bundles whose relative Euler characters $\chi(F_i,F_j)$
form an upper-triangular matrix.

%

The exceptional collections have been worked out for many interesting singularities.
For the purpose of this paper
all that we are going to need is the charge vector or Chern character $\ch{F_i}$ of the branes
in the exceptional collection. The Chern character of a sheaf tells us the
rank, the fluxes, and the instanton number, in other words it tells
us the effective (D7,D5,D3) wrapping numbers of the fractional
brane.

Thus to an exceptional collection we can associate a quiver diagram.
Each sheaf in the collection corresponds to an irreducible
fractional brane, and thus to a node. The net number of chiral
fields between two nodes is simply the net intersection number of
the cycles that the fractional brane wraps. We can put this in the
form of a matrix, the adjacency matrix of the quiver. In the case of
collapsed 4-cycles this is just the anti-symmetrization
$\chi_{-}(F_i,F_j)$ of the upper-triangular matrix of the
collection. The non-chiral matter can be obtained by a slightly more
refined cohomology computation.

As mentioned we can also reconstruct other F-term data such as the
superpotential. The physicists method is to compute some correlation
functions of the chiral fields. The mathematicians method is to
first compute the dual exceptional collection, whose relative Euler
characters are given by the inverse of the above mentioned
upper-triangular matrix. The superpotential now follows from the
relations in the path algebra of the dual collection.

%
 \begin{figure}[th]
\begin{center}
            \scalebox{.5}{
               \includegraphics[width=\textwidth]{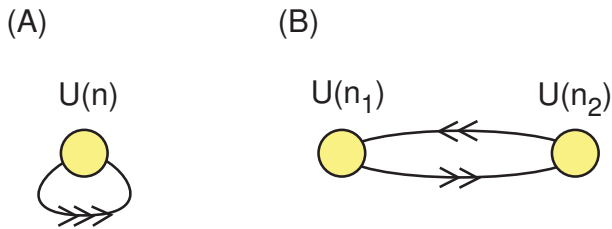}
               }
\end{center}
\vspace{-.5cm} \caption{ \it (A): Quiver for $N=4$ Yang-Mills theory. (B): Quiver
for the conifold.}\label{CanQuivers}
 \end{figure}
The superpotential encodes all of the complex geometry of the
Calabi-Yau singularity. This complex geometry is generically
non-commutative. Let us consider for example pure $N=4$ Yang-Mills
theory. Its quiver is a single node with 3 arrows back to itself.
The
superpotential is
\be W = {1\over 6}\,\epsilon_{ijk}\,  {\rm Tr}(Z^i Z^j Z^k) \qquad
i,j,k = 1,2,3. \ee
The $Z^i$'s are matrices after we assign gauge group ranks to the nodes, but
let us temporarily treat them as formal non-commuting variables.
The F-term equations then tell us that
\be \half \epsilon_{ijk}\, Z^i Z^j = 0 \ee
in other words the Calabi-Yau is a commutative ${\bf C}^3$. However
we may perturb the superpotential, for instance by adding mass terms
\be W = {1\over 6}\, \epsilon_{ijk}\, {\rm Tr}(Z^i Z^j Z^k) + \half
m_{kl}\, {\rm Tr}(Z^k Z^l) .\ee
The new F-term equations tell us that
\be \half \epsilon_{ijk} Z^i Z^j = m_{kl} Z^l \ee
In other words, we may deform ${\bf C}^3$ to a generic 3-dimensional Lie algebra.
This illustrates another important point that we also emphasized
in the introduction. The quiver gauge theory
itself is in fact the best {\it definition} of the local geometry.

As another example, let us consider the quiver for the conifold. It
has a superpotential
\be W = {\rm Tr}(A^i B^k A^j B^l)\, \epsilon_{ij} \epsilon_{kl},
\qquad i,j,k,l = 1,2. \ee
If we define
\be z_1 = A^1 B^1, \quad z_2 = A^1 B^2, \quad z_3 = A^2 B^1, \quad
z_4 = A^2 B^2 \ee
then the F-term equations tell us that
\be z_i z_j - z_j z_i = 0, \quad z_1 z_4 - z_2 z_3 = 0 . \ee
Superpotential deformations correspond to deformations of these
equations. For instance we could turn on mass terms
\be
W \to W + m_{ij} \, {\rm Tr}(A^i B^j). \ee
This leads to the relations
\be
\begin{array}{rclrcl}
  z_1 z_2 - z_2 z_1 & = & m_{21} z_1 + m_{22} z_2 \qquad & z_2 z_3 - z_3 z_2 & = & m_{22} z_4 - m_{11} z_1  \\
  z_3 z_1 - z_1 z_3 & = & m_{12} z_1 + m_{22} z_3 & z_2 z_4 - z_4 z_2  & = & m_{11} z_2 + m_{21} z_4 \\
  z_1 z_4 - z_4 z_1 & = & m_{12} z_2 - m_{21} z_3  & z_4 z_3 - z_3 z_4  & = & m_{11} z_3 + m_{12} z_4 \\
   &  &  & z_1 z_4 - z_2 z_3 & = & - m_{11} z_1 - m_{12} z_2
\end{array}
\ee
This `massive conifold' is the analogue of the $N=1^*$ deformation
of $N=4$ Yang-Mills theory. Actually this is only part of the story,
because the superpotential is modified quantum mechanically. In the
IR both gauge groups will confine and lead to glueball condensates.
Presumably this leads to a combination of a conifold transition and
a Myers effect.\footnote{In fact there are some natural conjectures
one can make because the vacua are largely constrained by the
representation theory of $SU(2) \times SU(2)$. Classically the
conifold has an $S^2$ at the bottom with a $B$-field through it, and
a transverse $S^3$. Then turning on the mass terms breaks the $SU(2)
\times SU(2)$ isometry, but for certain masses there is a linear
combination corresponding to some $S^2 \subset {\bf P}^1 \times {\bf
P}^1$ which is preserved. Eg. if we turn on $W \to W + m \,{\rm
Tr}(A^1 B^2 - A^2 B^1)$ then we would preserve the diagonal ${\bf
P}^1$, and vacua would be labelled by representations of the
diagional $SU(2)$. This presumably causes some of the branes to
expand to wrap the preserved  $S^2$ with a radius depending on $m$.
Turning on the glueball superpotential should lead to a conifold
transition. Now we should end up with a D5-brane, or in the S-dual
picture an NS5-brane wrapping the preserved $S^2$. Note that if we
take the diagonal $S^2$ to be preserved, then we seem to end up with
an NS5-brane wrapping $S^2 \subset S^3$ on the deformed conifold.
This would be a supersymmetric configuration but it is very
reminiscent of the KPV meta-stable vacuum \cite{Kachru:2002gs}. }
This theory exhibits many further interesting effects like
meta-stable vacua. Surprisingly it has received no attention in the
literature and we are further investigating it \cite{progress}.

More generally, we will be interested in adding irrelevant terms to
the superpotential. These clearly correspond to subleading complex
structure deformations of the singularity.

The physical intuition is that closed string modes are in one-to-one
correspondence with general gauge invariant deformations of the
quiver. For superpotential deformations this has been put on a firm
footing by Kontsevich \cite{Kontsevich}, who shows that
infinitesimal deformations of the ``derived category'' (i.e. single
trace superpotential deformations) correspond to observables in the
closed string B-model. In the context of mirror symmetry, the
significance of this statement is that together with a corresponding
statement for the A-model, it provides evidence for the
correspondence principle, i.e. the idea that classical mirror
symmetry can be recovered from homological mirror symmetry.

As we explained, our main interest will be in the Del Pezzo quivers.
The first five Del Pezzo quivers were found using orbifold and toric
techniques
\cite{Douglas:1996sw,Morrison:1998cs,beasley,Beasley:1999uz,Feng:2000mi}.
Some of these were rederived using exceptional collections in
\cite{Douglas:2000qw,Cachazo:2001sg}, and finally the remaining five
non-toric Del Pezzo quivers, including Del Pezzo 5 which will play a
central role in this paper, were found using exceptional collections
\cite{Wijnholt:2002qz}. We refer to
\cite{Wijnholt:2002qz,Wijnholt:2005mp} for more detailed reviews and
explicit computations. For other interesting works we refer to
\cite{Bergman:2005kv,Aspinwall:2004mb,Bridgeland:2005my,Herzog:2003dj,Herzog:2006bu}.

\newpage

\newsection{Orientifolding quivers}

Discussions of orientifolds and derived categories have recently
been given in the LG regime \cite{Hori:2006ic} and in the large
volume regime \cite{Diaconescu:2006id}. Here we describe
orientifolds in another regime, which is captured by quiver gauge
theories. Traditionally orientifolds of branes at singularities have
been derived by first specifying an orientifold action on the closed
string modes, and then finding the induced action on open string
modes. Here we start by specifying an orientifold action on the open
string modes. This simplifies the task of finding a brane
realization of a desired gauge theory, and at any rate the closed
string geometry can be reconstructed from the gauge theory.

\newsubsection{General discussion of orientifolding}

Perturbative string theory on IIb backgrounds has a number of
$Z_2$ symmetries. They include $(-1)^{F_L}$ and worldsheet parity
$P$. In addition, on a given background the theory may have an
additional $Z_2$ symmetry $\sigma$.

Given such a symmetry, we can construct a new perturbative string
background by gauging it. An orientifold projection is an orbifold
which involves $P$. In addition, we would like to preserve $N=1$
SUSY in four dimensions. Recall that in IIB string theory on a Calabi-Yau the
supercharges with positive 4D chirality are derived from the
currents
\be j^1_\alpha =  e^{-\varphi_L/2}\,S_{L\alpha}\,e^{{1\over 2}\int
J_L} , \qquad j^2_\alpha = e^{-\varphi_R/2}\,S_{R\alpha}\,e^{{1\over
2}\int J_R} , \ee
where in the large volume limit
\be e^{\int J_L} = \Omega^{(3,0)}_{ijk} \psi_L^i \psi_L^j \psi_L^k,
\qquad e^{\int J_R} = \Omega^{(3,0)}_{ijk} \psi_R^i \psi_R^j
\psi_R^k. \ee
We have used the conventional notations
for the bosonized superghost, 4D spin fields and wordsheet fermions. In type IIa,
the second current would have been proportional to the square root of
$\bar{\Omega}^{(0,3)}_{\bar{i}
\bar{j}\bar{k}}\psi^{\bar{i}}_R \psi^{\bar{j}}_R \psi^{\bar{k}}_R$,
because the second spinor must have negative 10D chirality.
In order to preserve SUSY there must be a linear combination that is preserved.
If we do not include $(-1)^{F_L}$, then
\be
Q^1_\alpha + Q^2_\alpha
\ee
is preserved under orientifolding, provided $\sigma$ is a symmetry
of the internal CFT  that maps $\Omega^{(3,0)} \to \Omega^{(3,0)}$.
If we instead include $(-1)^{F_L}$ in the orientifolding, then
\be
Q^1_\alpha -i Q^2_\alpha
\ee
is preserved under orientifolding, provided $\sigma$ maps
$\Omega^{(3,0)} \to
-\Omega^{(3,0)}$.

Parity exchanges the Chan-Paton factors at the ends of an open
string, and acts as $-1$ on massless open string modes, so it maps
gauge fields and chiral fields to minus their transpose. We are
interested in local orientifold models, so we will be looking for
symmetries of the quiver of irreducible branes which map a gauge
field at node $i$ to minus the transpose of the gauge field at some
node $j$, and map any chiral field $X$ to the transpose $Y^T$ of
some other chiral field, possibly up to an additional gauge
transformation which we call $\gamma$. We denote this as $i
\leftrightarrow j^*$. In our decoupling limit, finding such a
symmetry is sufficient, because the irreducible branes generate all
other branes and closed strings ought to be recovered from open
strings. In particular we can read of the local geometry from the
gauge invariant operators and their relations.

We will assume canonical kinetic terms for the chiral fields, so we
can actually map $X \to e^{i\varphi} Y^T$ for some phase $\varphi$.
If there are multiple arrows between two nodes, we can upgrade
the map to a unitary matrix. In order to preserve SUSY, the orientifold
action has to leave the superspace coordinate $\theta$ invariant,
and hence it will also have to leave the superpotential invariant.
This may lead to correlations between the $SO$ and $Sp$ projections
on different nodes, and symmetric or anti-symmetric projections on
chiral fields that are mapped to themselves.

One should keep in mind that a non-anomalous quiver theory may
become anomalous after projection if the ranks of the gauge groups
are not adjusted. The orientifold may project out more of the
positive than of the negative contributions to an anomaly. This is
generically the case if the projected theory contains symmetric and
anti-symmetric tensor matter. From the geometric point of view, this
is because an orientifold plane may give additional tadpole
contributions, which need to be cancelled by adding additional
branes, i.e. adjusting the ranks of the gauge groups.

We can also understand how the orientifold acts on closed string
modes. The modes that are kept are simply the closed string modes on
the cover that can be used to deform the orientifolded theory. To
preserve SUSY, the orientifold action maps $\tau_i \to \tau_{j^*}$
and $L_i \to - L_{j^*}$, where $\tau$ denotes the complexified gauge
coupling and $L$ denotes the linear multiplet containing the FI
parameter and the St\"uckelberg 2-form field.

\newsubsection{Examples}

\noindent {\it Orientifold of ${\bf P}^2$}

%
 \begin{figure}[th]
\begin{center}
            \scalebox{.6}{
               \includegraphics[width=\textwidth]{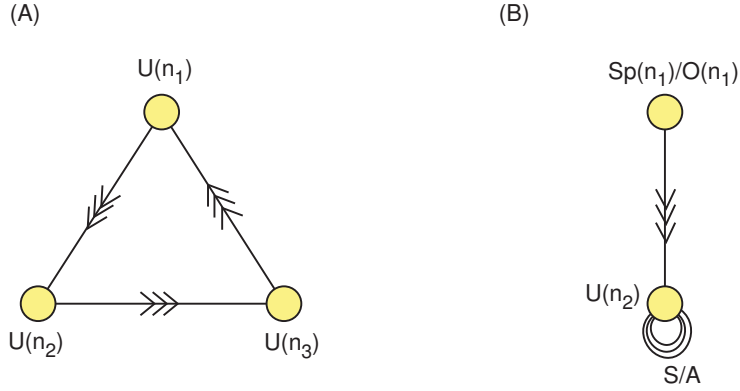}
               }
\end{center}
\vspace{-.5cm} \caption{ \it (A) Quiver for ${\bf P}^2$. (B)
Orientifold.}\label{P2orientifold}
 \end{figure}

The simplest case to understand is the Calabi-Yau cone over ${\bf
P}^2$, which is identical to the orbifold singularity ${\bf
C}^3/Z_3$. This orientifold is already known in the literature \cite{Lykken:1998ec},
but we will use a slightly more geometric perspective. We denote the
hyperplane class by $H$. The quiver is given in figure
\ref{P2orientifold}A and may be obtained from a set of fractional
branes with the following (D7,D5,D3) wrapping numbers:
\be 1.\ (1,0,0) \quad 2.\ -(2,H,-\half) \quad 3.\ (1,H,\half) . \ee
We consider the following symmetry:
\be 1 \leftrightarrow 1^*, \quad 2 \leftrightarrow 3^*. \ee
We may take an $Sp$- or $SO$-projection. Assuming the usual orbifold
superpotential, the matter between nodes 2 and 3 projects to a
conjugate symmetric tensor (for $Sp$) or a conjugate anti-symmetric
tensor (for $SO$). The orientifolded quiver is given in figure
\ref{P2orientifold}B.

We expect an orientifold plane which coincides with the fractional
brane on node 1, i.e. it is an O7-plane wrapped on the vanishing Del
Pezzo. Anomaly cancellation implies that $n_1 = n_2 + 4$ for
symmetric tensor matter, and $n_1 = n_2 - 4$ for anti-symmetric
tensor matter. In order to cancel the flux through the hyperplane
class, we take the charge vector of the O7-plane to be $(\mp
4,0,0)$. We don't guarantee however that there are no further
O3-plane charges.

In the geometric regime the net number of symmetric and
anti-symmetric matter is given by \cite{Blumenhagen:2002wn}
\ba \# \ {\rm sym} \is \half I_{nn^*} + {1\over 8} I_{nO} \eol \#\
{\rm asym} \is \half I_{nn^*} - {1\over 8} I_{nO} \ea
where $I_{ij}$ is the intersection form of the Calabi-Yau. We expect
this formula also for small volume, and indeed it agrees with the
spectrum above. However it is not always clear what charges we
should assign to an orientifold plane. The guiding principle is that
we get a sensible gauge theory in which all anomalies are cancelled,
and from that we may try to reconstruct the orientifold plane.

The proposed O7-plane is not the fixed locus of any $Z_2$ after
blowing up, so we believe that the large volume limit is projected
out. There are other ways to see this. The definition of the
orientifold which produces this quiver involves a symmetry which
interchanges oppositely twisted sectors, which is not available
after blowing up. There is no symmetry that maps a rank 2 bundle to
a rank 1 bundle in the geometric regime. And the orientifold imposes
relations on the gauge couplings of nodes $2$ and $3$, which in turn
freezes the K\"ahler modulus.

The case of the $SO$ projection with $n_2 = 5$ gives us a simple
3-generation $SU(5)$ GUT with a $\bar{5}$ and a $10$ from the
anti-symmetric \cite{Lykken:1998ec}. There are no Higgses, though
they could presumably be generated by first increasing the ranks and
then Higgsing. Of course there would be well-known problems with
getting the ${5}\times 10\times 10$ Yukawa's. This model also
exhibits dynamical SUSY breaking \cite{Lykken:1998ec}, but with a
runaway behaviour.

\bigskip

\noindent {\it Orientifold of ${\bf P}^1 \times {\bf P}^1$}

%
 \begin{figure}[th]
\begin{center}
            \scalebox{.6}{
               \includegraphics[width=\textwidth]{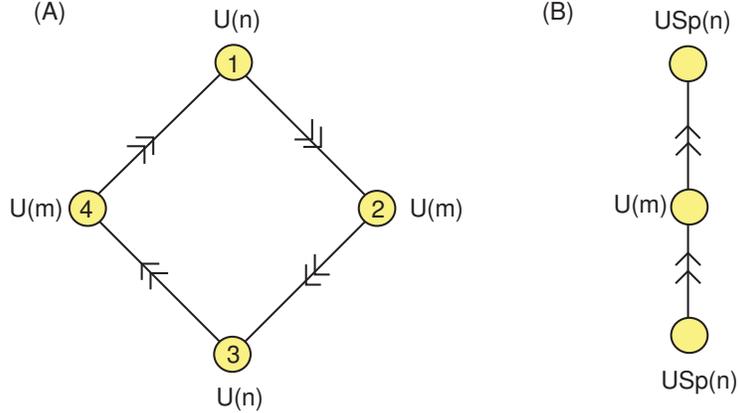}
               }
\end{center}
\vspace{-.5cm} \caption{ \it (A) Quiver for ${\bf P}^1 \times {\bf
P}^1$. (B) Orientifold.}\label{P1xP1}
 \end{figure}

The next interesting case is ${\bf P}^1 \times {\bf P}^1$. This
example is very similar to the ${\bf C}^3/Z_4$ orbifold singularity,
to which it is related by turning on masses for the non-chiral
fields and flowing to the IR. We denote the first ${\bf P}^1$ by
$H_1$ and the second by $H_2$. Their intersection numbers, when
restricted to ${\bf P}^1 \times {\bf P}^1$, are
\be
 H_1 \cdot H_1 = H_2 \cdot H_2 = 0, \quad H_1 \cdot H_2 = 1.
 \ee
The quiver is shown in figure \ref{P1xP1} and can be obtained from
an exceptional collection with the following $(D7,D5,D3)$ wrapping
numbers:
\be
 1.\ (1,0,0)  \qquad 2.\ -(1,H_1,0) \qquad 3.\ -(1,H_2\!-\! H_1,1)
\qquad 4.\ (1,H_2,0) \ee
%
%
%
We are interested in the following orientifold action:
\be\label{P1xP1Z2}
1\leftrightarrow 1^* \qquad 2\leftrightarrow 4^* \qquad 3 \leftrightarrow 3^*
\ee
and
\be\label{P1xP1proj}
X_{12}^T \gamma_1 = X_{41}, \qquad Y_{12}^T \gamma_1 = Y_{41}, \qquad
\gamma_3^{-1} X_{23}^T = X_{34}, \qquad \gamma_3^{-1} Y_{23}^T = Y_{34}.
\ee
As usual we have
\be
\gamma_1^T = s(1)\cdot \gamma_1, \qquad \gamma_3^T = s(3)\cdot \gamma_3
\ee
with $s = \pm 1$, in order for the action on the gauge fields to be
an involution. The superpotential of the commutative ${\bf P}^1
\times {\bf P}^1$ quiver is
\be W =
 X_{12} X_{23} X_{34} X_{41} - X_{12} Y_{23} X_{34} Y_{41}
+ Y_{12} Y_{23} Y_{34} Y_{41} - Y_{12} X_{23} Y_{34} X_{41} . \ee
In
order for this particular superpotential to be invariant, we also
need
\be
s(1)\cdot s(3) = +1
\ee
so we can have an $SO/SO$ or an $Sp/Sp$ projection. More generally
we could work the other way around. We first decide on the
projections that we would like to have, and then we write down the
most general superpotential compatible with those projections.

The orientifold locus appears to consist of the union of two
O7-planes, with wrapping numbers $4(1,0,0)$ and $-4(1,H_2\!-\!
H_1,1)$. This is not the fixed locus of any $Z_2$ symmetry after
blowing up, so the large volume limit is projected out.

\bigskip
\noindent {\it Orientifold of Del Pezzo 5}

\begin{figure}[th]
\begin{center}
            \scalebox{.3}{
               \includegraphics[width=\textwidth]{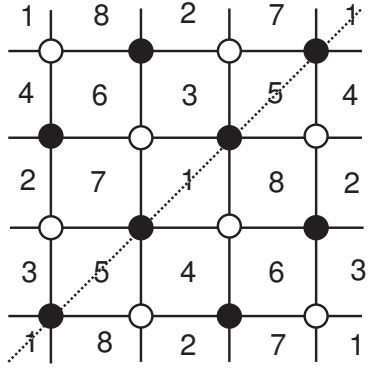}
               }
\end{center}
\vspace{-.5cm} \caption{ \it  Dimer graph/brane box picture for a
toric degeneration of Del Pezzo 5.}\label{DP5dimer}
 \end{figure}

Now we come to the main case of interest, the Del Pezzo 5
singularity. The Del Pezzo 5 surface is a ${\bf P}^2$ blown up at
five generic points. As a basis for the 2-cycles we use the
hyperplane class $H$ and the exceptional curves created by the
blow-ups, $E_i,\ i = 1 \ldots 5$, with the intersections
\be H\cdot H = 1, \quad H \cdot E_i = 0, \quad E_i \cdot E_j = -
\delta_{ij}. \ee
We can construct the DP5 quiver from a collection of line bundles
with the following charge vectors:
\be
\begin{array}{ll}
  1.\ (1,H\!-\! E_1,0) & 3. -\!(1,2H\! -\! E_1\! - \! E_2\! -\! E_4,\half) \\
  2.\ (1,H\!-\!E_2,0)\quad   & 4. -\!(1,2H\! -\! E_1\! -\! E_2\! -\! E_5,\half)\  \\
                        &           \\
     5. -\!(1,H\!-\! E_3,0) &        7.\ (1,H,\half) \\
      6. -\!(1,E_4\! +\! E_5,-1)\quad   & 8.\ (1,2H \!-\! E_1\! -\! E_2\! - \!E_3,\half)
\end{array}
\ee
%
%
This singularity has a well-known toric limit which is the $Z_2
\times Z_2$ orbifold of the conifold. This limit will not have any
special significance for us, but we point it out because it is
perhaps more familiar to the reader. In the toric limit the
superpotential can be graphically represented through a dimer
diagram (we refer to \cite{Franco:2005rj} for dimer rules). Since
orientifolding leaves the superpotential invariant, it must
correspond to a reflection or $180^0$ degree rotation of the dimer.
The toric superpotential is read off to be:
\ba W \is X_{13} X_{35} X_{58} X_{81} - X_{14} X_{46} X_{68} X_{81}
+ X_{14} X_{45} X_{57} X_{71} - X_{13} X_{36} X_{67} X_{71} \eol & &
+X_{24} X_{46} X_{67} X_{72} - X_{23} X_{35} X_{57} X_{72} + X_{23}
X_{36} X_{68} X_{82} - X_{24} X_{45} X_{58} X_{82}. \ea
and it is invariant under the reflection in the axis indicated in
figure \ref{DP5dimer}.
%
%
 \begin{figure}[th]
\begin{center}
            \scalebox{.8}{
               \includegraphics[width=\textwidth]{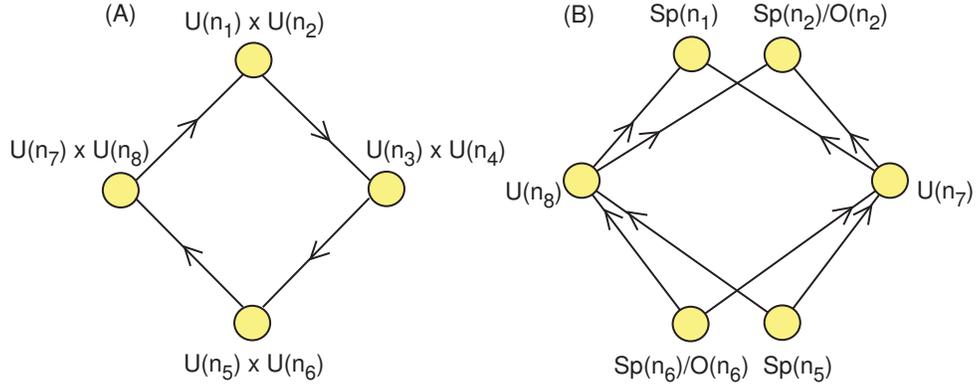}
               }
\end{center}
\vspace{-.5cm} \caption{ \it (A): The Del Pezzo 5 quiver. (B):
Orientifold associated to the toric $Z_2$
symmetry.}\label{orientiDP5}
 \end{figure}

%

We are interested in the following orientifold action:
\be\label{DP5Z2}
1 \leftrightarrow 1^*, \quad 2 \leftrightarrow 2^*, \quad 3 \leftrightarrow 8^*,
\quad 4 \leftrightarrow7^*, \quad 5 \leftrightarrow 5^*, \quad 6 \leftrightarrow 6^*.
\ee
The action on the fields is
\be\label{DP5proj}
\begin{array}{cccc}
  X_{13}^T \gamma_1 = X_{81}  & X_{14}^T \gamma_1 = a \,X_{71}
  & X_{23}^T\gamma_2 = b X_{82} & X_{24}^T \gamma_2 = X_{72} \\
  \gamma_5^{-1} X_{35}^T = X_{58} & \gamma_5^{-1} X_{45}^T = X_{57}
  & \gamma_6^{-1} X_{36}^T= X_{68} & \gamma_6^{-1} X_{46}^T = X_{67}
\end{array}
\ee
where $a$ and $b$ are phases. If we insist on taking the toric
superpotential, then invariance of the superpotential implies
\be
s(1) s(5) = s(1) s(6) a = s(1) s(6) a^{-1} = s(2) s(5)b^{-1} = s(2) s(5) b = s(2) s(6) = 1
\ee
Hence with this superpotential, the projections on nodes 1 and 5,
and the projections on nodes 2 and 6 are always the same, but other
than that it is free to be chosen. In particular the all $Sp$
projection that we will use for our first construction is actually
realized in the toric limit, and can be seen for instance in the
dimer description. However we will consider generic superpotentials
compatible with the projection.

The orientifold locus should consist of the union of four O7-planes,
coinciding with the fractional branes of nodes $1,2,5$ and $6$.

\bigskip
\noindent {\it Another orientifold of Del Pezzo 5}

We consider the same Del Pezzo 5 quiver, but with an alternative
orientifold action
\be\label{altDP5} 1 \to 1^*,\qquad  2 \to 2^*, \qquad 3 \to
8^*,\qquad 4 \to 7^*, \qquad 5 \to 6^*. \ee
The conditions on the fields are
\be\label{altDP5orient} X_{13}^T \gamma_1 = X_{81}, X_{14}^T
\gamma_2 = X_{71}, X_{23}^T \gamma_2 = X_{82}, X_{24}^T \gamma_2 =
X_{72},\ee \be X_{35}^T = X_{68}, X_{36}^T = X_{58}, X_{45}^T =
X_{67}, X_{46}^T = X_{57} . \ee
We take the $Sp/Sp$ projection on nodes 1 and 2. Presumably there
are two O7-planes, coinciding with the fractional branes on nodes
$1$ and $2$.

 \begin{figure}[th]
\begin{center}
            \scalebox{.4}{
               \includegraphics[width=\textwidth]{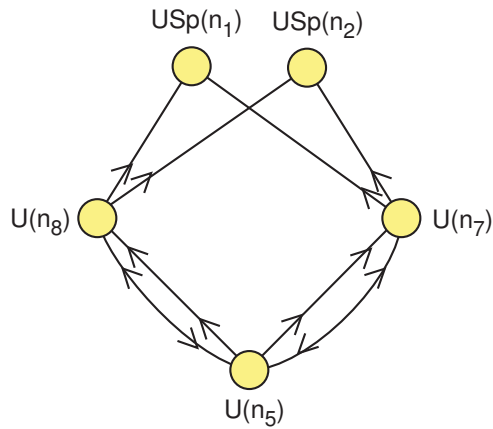}
               }
\end{center}
\vspace{-.5cm} \caption{ \it Another interesting orientifold of Del
Pezzo 5.}\label{AltOrientiDP5}
 \end{figure}

\bigskip
\noindent
{\it Del Pezzo 7}

Several other Del Pezzo quivers could be used for MSSM constructions.
We briefly mention a quiver for Del Pezzo 7. The vanishing homology classes
again consist of the class of the 4-cycle, 0-cycle, the hyperplane
class $H$, and the exceptional curves $E_i,i = 1, \ldots, 7$, with
the intersection numbers
\be H\cdot H = 1, \quad H \cdot E_i = 0, \quad E_i \cdot E_j = -
\delta_{ij}. \ee
An exceptional collection is given by
%
\be
\begin{array}{llll}
  1.\ \ (2,H,-\half) & 2.\ -(1,H\!-\!E_5,0) \\
            & 3.\ -(1,H\!-\!E_6,0) \\
            & 4.\ -(1,H\!-\!E_7,0) \\
            & 5.\ -(1,3H\!-\!\sum_i E_i,1) \\
            &                              \\
 6.\  -(2,E_5\!+\!E_6\!+\!E_7,-{3\over 2})\qquad  & 7.\ \ (1,H\!-\!E_1,0) \\
  & 8.\ \  (1,H\!-\!E_2,0) \\
  & 9.\ \ (1,H\!-\!E_3,0) \\
    & 10.\,(1,H\!-\!E_4,0) \\
\end{array}
\ee
One way to orientifold this quiver is by reflecting in the axis
through nodes $1$ and $6$.

 \begin{figure}[th]
\begin{center}
            \scalebox{.4}{
               \includegraphics[width=\textwidth]{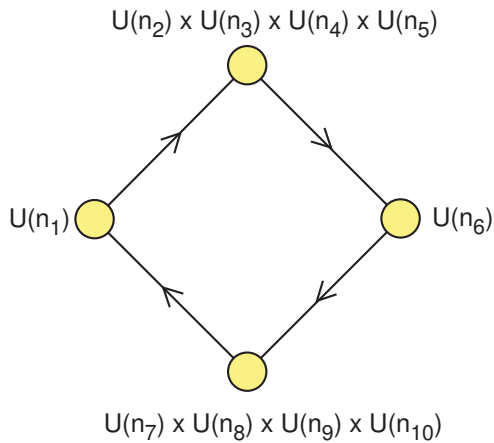}
               }
\end{center}
\vspace{-.5cm} \caption{ \it Quiver for Del Pezzo
7.}\label{DP7Quiver}
 \end{figure}

\newpage

\newsection{The Higgsing procedure}

\newsubsection{Model I}
\label{ModelIsection}

Now we would like to engineer the MSSM quivers we have discussed. We
take the quiver in figure \ref{orientiDP5} with an $Sp/Sp/Sp/Sp$
projection. As far as the chiral field content goes, this contains
the MSSM with one generation of quarks and leptons. In order to
increase the number of generations, we have to create some
non-trivial bound states of fractional branes.

Let us try to give a rather imprecise but intuitive geometric
picture of our procedure (which we will then promptly abandon in
favour of more precise statements). Each fractional brane
corresponds to a line bundle on the Del Pezzo surface, i.e. a
non-trivial $U(1)$ gauge field configuration. Roughly we want to
take three identical fractional branes (which corresponds to a
$U(1)^3$ field configuration with $U(1)$ holonomy) and add some
instantons to get a field configuration with $U(3)$ holonomy on the
Del Pezzo. Recall that the 4D gauge symmetry is the subgroup of the
gauge group on the brane that commutes with the holonomy. This new
fractional brane then has the same intersection numbers as the
original brane, times a factor of three. Any moduli of the new
fractional brane can be lifted by turning on suitable $B$-fields.

The more pedestrian and precise statements are that we first
increase the ranks of the gauge groups and then turn on suitable
VEVs in order to get to the quiver we want. We claim that there
exists a bound state $F_b$ with the following charge vector:
\ba \ch{F_b} &=& \sum_{i} n_i \, \ch{F_i} = -(3,2H\!-\! E_2\!
-\!E_3\!+\!E_4\!+\!E_5,-2),\eol
 \vec{n} &=& \{ 1,0,1,1,2,2,1,1 \}.
\ea
%
%
To see this, let's fix the gamma matrices to be
\be
\gamma_1 = i\sigma_2 =
\left(
  \begin{array}{cc}
    0 & 1 \\
    -1 & 0 \\
  \end{array}
\right),
\quad
\gamma_5 = \gamma_6 =
\left(
  \begin{array}{cc}
    i\sigma_2 & 0 \\
    0 & i\sigma_2 \\
  \end{array}
\right),
\ee
and consider the following VEVs:
\be X_{13} =
\left(
  \begin{array}{c}
    b_1 \\
  \end{array}
\right)\!,
\ X_{35} =
\left(
  \begin{array}{cc}
    b_2 & 0 \\
  \end{array}
\right)\!,
\ X_{36} =
\left(
  \begin{array}{cc}
    b_3 & 0 \\
  \end{array}
\right)\!, \ee
\be X_{14} =
\left(
  \begin{array}{c}
    d_1 \\
  \end{array}
\right)\!,
\ X_{45} =
\left(
  \begin{array}{cc}
    0 & d_2 \\
  \end{array}
\right)\!,
\ X_{46} =
\left(
  \begin{array}{cc}
    0 & d_3 \\
  \end{array}
\right)\!, \ee
with the remaining fields determined by the orientifold conditions.
The entries here are $ 2\times 2$ matrices. These VEVs break the gauge
symmetry to $Sp(2)$, so the fractional brane $F_b$ will come with an
$Sp$-projection. The D-term equations may be satisfied by taking
\be
{1\over 5} b_1 = {1\over 4} b_2 = {1\over 3} b_3 =
\left(
  \begin{array}{cc}
    \chi_1 & 0 \\
    0 & \chi_1 \\
  \end{array}
\right), \qquad
{1\over 5} d_1 = {1\over 4} d_2 = {1\over 3} d_3 =
\left(
  \begin{array}{cc}
    \chi_2 & 0 \\
    0 & \chi_2 \\
  \end{array}
\right).
\ee
Clearly we need both quartic and octic terms in the superpotential,
in order to get mass terms for the  adjoints
associated with a rescaling of the VEVs. With such a superpotential,
tuned so
that the F-term equations are satisfied and so that the orientifold
symmetry is preserved, but otherwise generic, we find that our bound
state is rigid, i.e. it has no massless adjoints.

We also claim that there exists a bound state $F_a$ with the
following charge vector:\footnote{ This charge vector is probably
not the Chern character of a sheaf; instead it should be interpreted
as a bound state of branes and ghost branes in the large volume
limit \cite{Wijnholt:2005mp}.}
\ba \ch{F_a} &=& \sum_{i} n_i \, \ch{F_i}=
(3,2H\! -\! E_1\! -\! E_2\! +\! E_4\! +\! E_5,0), \eol
 \vec{n} &=& \{
2,2,1,1,1,0,1,1 \}
\ea
%
%
This is very similar to $F_b$ so we do not need to repeat the
analysis. The fractional brane $F_a$ also inherits an $Sp$
projection.

\begin{figure}[th]
\begin{center}
            \scalebox{.5}{
               \includegraphics[width=\textwidth]{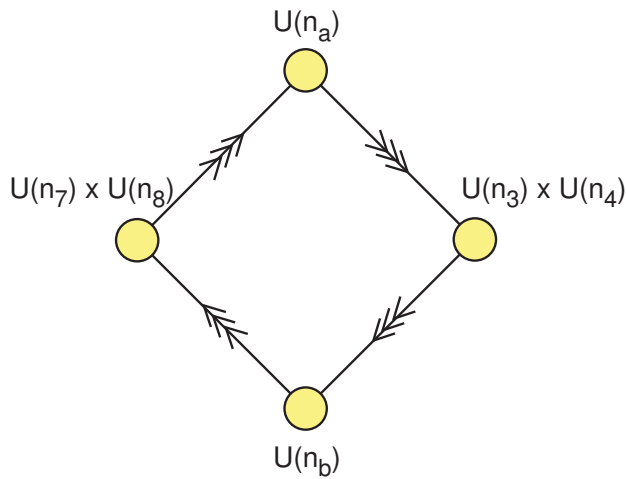}
               }
\end{center}
\vspace{-.5cm} \caption{ \it Intermediate
quiver.}\label{Model1preQuiver}
 \end{figure}

Next we compute the massless fields in the quiver for $\{ 2 F_a,3
F_3,F_4,2F_b,F_7,3F_8 \}$. For a generic superpotential (apart from
the conditions mentioned above) we found that the spectrum is
completely chiral, as shown in figure \ref{Model1preQuiver}. This
quiver is a Higgsed version of the original Del Pezzo quiver. It
inherits  the following orientifold projection:
\be X^{iT}_{a3} \gamma_a = X^i_{8a} , \quad
 X_{a4}^{iT} \gamma_a =  X_{7a}^i, \quad
\gamma_b^{-1} X_{3b}^{iT} = X^i_{b8} , \quad
 \gamma_b^{-1} X_{4b}^{iT} = X^i_{b7}, \quad i=1,2,3.
\ee
Moreover, we expect to be able to get a generic superpotential for
this quiver provided we included sufficiently many higher order
terms in the original Del Pezzo quiver. Checking we can get generic
4th and 8th order terms is computationally too intensive, so we will
assume it from now on.

This is almost what we want. After orientifolding we get all the
chiral fields of the MSSM. However, we also want some non-chiral
matter: the conventional Higgses $H_u,H_d$ and the additional Higgs
fields which break $Sp(2)_R \times U(1)_L \to U(1)$. This cannot be
obtained by tuning the original bound state/superpotential, because
the candidate non-chiral fields are in fact eaten by gauge bosons.
So we create a new quiver with the same chiral matter content, but
with more candidate non-chiral fields.\footnote{Alternatively, we
could use more complicated bound states from the beginning, but then
we would have to work with superpotential terms of order 12 or
higher in order to lift the excess non-chiral matter.} To do this we
replace $F_b$ by the bound state $F_d$ with charge vector
\ba {\rm ch}(F_d) &=& {\rm ch}(F_a) + {\rm ch}(F_3) + 2 \, {\rm ch}(F_b)
+ {\rm ch}(F_8) \eol
&=& -(3,2H+E_1-E_2-E_3+E_5,-4)
\ea
This can be done for instance by turning on VEVs of the following
form:
\be X_{a4}^1 =
\left(
  \begin{array}{c}
    s_1 \\
  \end{array}
\right) \quad
X_{4b}^1 =
\left(
  \begin{array}{cc}
    s_2 & 0 \\
  \end{array}
\right)
\quad
X_{4b}^2 =
\left(
  \begin{array}{cc}
    0 & s_3 \\
  \end{array}
\right) \ee
and the remaining non-zero VEVs fixed by the orientifold conditions.
We can satisfy the D-terms by setting
\be
{1\over 5} s_1 = {1\over 4} s_2 = {1\over 3} s_3 =
\left(
  \begin{array}{cc}
    \phi & 0 \\
    0 & \phi \\
  \end{array}
\right).
\ee
%
In order for this to satisfy the F-term equations, and to get the
desired massless non-chiral matter, we have to impose some
restrictions on the superpotential. If we use both quartic and octic
terms, one can lift all the non-chiral matter and the quiver
generated by $\{ 2 F_a,3 F_3,F_4,2 F_d,F_7,3 F_8 \}$ is the same as
in figure \ref{Model1preQuiver} again. However now there are two
non-chiral pairs between $F_a$ and $F_d$ and two non-chiral pairs
between $F_d$ and $F_4/F_7$. We have checked that the superpotential
can be tuned so that one of each of these pairs becomes light, and
so we end up with the required quiver in figure \ref{CoverQuivers}A.

\newsubsection{Pati-Salam}

The quiver we have obtained above also gives a three generation
SUSY Pati-Salam
model, by changing the ranks of the gauge groups ($U(3)\to U(0)$, and
$U(1) \to U(4)$).

\begin{figure}[th]
\begin{center}
            \scalebox{.25}{
               \includegraphics[width=\textwidth]{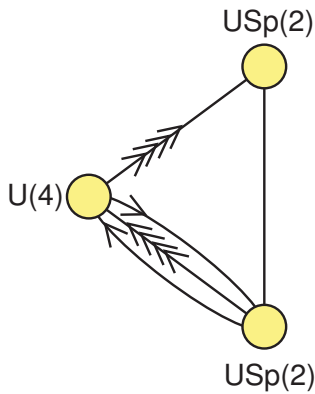}
               }
\end{center}
\vspace{-.5cm} \caption{ \it A four generation Pati-Salam model from
${\bf P}^1 \times {\bf P}^1$.}\label{4genPatiSalam}
 \end{figure}

Very similar tricks may also be applied to the ${\bf P}^1 \times
{\bf P}^1$ quiver to construct Pati-Salam models, though in that
case the number of generations will be even. Let us show how we can
obtain a four generation Pati-Salam model. We define the gamma matrices as
\be
\gamma_1 = \left(
\begin{array}{cc}
 0 & 1   \\
 -1 & 0  \\
 \end{array}
\right), \qquad
\gamma_3 =  \left(
               \begin{array}{cc}
                 0 & {\bf 1}_{3\times 3}   \\
                 -{\bf 1}_{3\times 3} & 0  \\
               \end{array}
             \right),
\ee
Then we construct a bound
state $F_b$ with charge vector
\be {\rm ch}(F_b) = \ch{ F_1} + 2 \ch{F_2} + 3\ch{ F_3} + 2 \ch{F_4} = -(2,H_2\! -\!
H_1,3) \ee
by turning on the following VEVs:
\be X_{12} = \left(
               \begin{array}{cccc}
                 a_1 & 0 & 0 & 0 \\
                 0 & 0 & a_1 & 0 \\
               \end{array}
             \right),
             \quad
 Y_{12} = \left(
               \begin{array}{cccc}
                 0 & a_1 & 0 & 0 \\
                 0 & 0 & 0 & a_1 \\
               \end{array}
             \right)
             \quad,
 \ee
\be
 X_{23}= \left(
           \begin{array}{cccccc}
             a_2 & 0 & 0 & 0 & 0 & 0 \\
             0 & 0 & a_3 & 0 & 0 & 0 \\
             0 & 0 & 0 & a_2 & 0 & 0 \\
             0 & 0 & 0 & 0 & 0 & a_3 \\
           \end{array}
         \right),
\quad Y_{23}= \left(
           \begin{array}{cccccc}
             0 & 0 & 0 & 0 & 0 & 0 \\
             0 & a_4 & 0 & 0 & 0 & 0 \\
             0 & 0 & 0 & 0 & 0 & 0 \\
             0 & 0 & 0 & 0 & a_4 & 0 \\
           \end{array}
         \right)
\ee
with the remaining VEVs determined by the orientifold symmetry. The
D-terms are satisfied if we pick
\be a_1 = a_2 = 5\psi, \quad a_3 = 4\psi, \quad a_4 = 3\psi. \ee
Similarly we construct a bound state $F_a$ with charge vector
\be {\rm ch}(F_a) = 3 \ch{F_1} + 2 \ch{F_2} + \ch{ F_3} + 2 \ch{F_4} = (2,H_2\! - \!
H_1,-1). \ee
Then by orientifolding the quiver generated by $\{ 2 F_a,4 F_2,2
F_b,4 F_4 \}$ we get the Pati-Salam quiver with four generations in
figure \ref{4genPatiSalam}. It's expected all excess non-chiral
matter can be lifted by an induced superpotential, but we did not
try very hard to do it explicitly in this case. The total
configuration has net wrapping number $(0,4H_2-4H_1,-8)$.

\newsubsection{Model II}
\label{ModelIIsection}

We now consider an alternative construction, in which $U(1)_{B-L}$
can get a mass through the St\"uckelberg mechanism (depending on the
UV completion \cite{Buican:2006sn}).

The orientifold projection is as in (\ref{altDP5}). We would like to
form the following bound states:
\ba
{\rm ch} (F_a) &=& \sum_{i} n_i\,  \ch{F_i} = (3,2H\!-\! 2E_1\!
+\! E_2\! -\! E_3\! +\! E_4\! +\! E_5,1), \eol
\vec{n}_a &=&
\{4,1,2,2,1,1,2,2\} \eol
{\rm ch} (F_b) &=& \sum_{i} n_i\,  \ch{F_i} = -(3,\! - \! E_1 \! -
\! 2 E_2 \! + \! 3 E_3 \! +\! 3 E_4\! +\! 3 E_5, 6), \eol
 \vec{n}_b &=&
\{2,1,3,3,0,6,3,3\} \eol
{\rm ch} (F_{b'}) &=& \sum_{i} n_i\,  \ch{F_i} = -(3,6H \!-\!
E_1\! -\! 2E_2\! -\! 3E_3\! -\! 3E_4\! -\! 3E_5,0), \eol
 \vec{n}_{b'} &=&
\{2,1,3,3,6,0,3,3\} \ea
%
For $2 F_a$ we suggest the following VEVs:
\be
X_{13} =
\left(
  \begin{array}{cccc}
    a_1 & 0 & 0 & 0 \\
    0 & a_2 & 0 & 0 \\
    0 & 0  & 0 & 0 \\
    0 & 0  & 0 & 0 \\
    0 & 0  & a_1 & 0 \\
    0 & 0 & 0 & a_2 \\
    0 & 0  & 0 & 0 \\
    0 & 0  & 0 & 0 \\
  \end{array}
\right), \quad
X_{14} =
\left(
  \begin{array}{cccc}
    0  & 0 & 0 & 0 \\
    0 & 0 & 0 & 0 \\
    a_2 & 0  & 0 & 0 \\
    0 & a_1  & 0 & 0 \\
    0 & 0  & 0 & 0 \\
    0 & 0 & 0 & 0 \\
    0 & 0  & a_2 & 0 \\
    0 & 0  & 0 & a_1 \\
  \end{array}
\right), \ee
\be
 X_{23} = \left(
  \begin{array}{cccc}
    a_3 & 0 & 0 & 0 \\
    0 & 0 & a_3 & 0 \\
  \end{array}
\right), \quad
X_{24} =
\left(
  \begin{array}{cccc}
   0 & a_3 & 0 & 0 \\
    0 & 0 & 0 & a_3 \\
  \end{array}
\right),
\ee
\be
X_{35} =
\left(
  \begin{array}{cc}
    a_4 & 0\\
    0 & 0 \\
    0 & a_4 \\
    0 & 0 \\
  \end{array}
\right),
\quad
X_{36} =
\left(
  \begin{array}{cc}
    0 & 0 \\
    0 & a_5 \\
    0 & 0 \\
    a_5 & 0 \\
  \end{array}
\right),
\quad
X_{45} =
\left(
  \begin{array}{cc}
    0 & 0 \\
    a_5 & 0 \\
    0 & 0 \\
    0 & a_5 \\
  \end{array}
\right),
\quad
X_{46} =
\left(
  \begin{array}{cc}
    a_4 & 0 \\
    0 & 0 \\
    0 & a_4 \\
    0 & 0 \\
  \end{array}
\right).
\ee
The remaining fields are determined by the orientifold conditions
(\ref{altDP5orient}). We took the gamma matrices to be
\be
\gamma_1 =
\left(
  \begin{array}{cc}
     0  & {\bf 1}_{4\times 4} \\
    -{\bf 1}_{4\times 4} & 0 \\
  \end{array}
\right), \qquad
\gamma_2 =
\left(
  \begin{array}{cc}
     0  & 1 \\
    -1& 0 \\
  \end{array}
\right).
\ee
The D-term equations reduce to
\be |a_1|^2 + |a_3|^2 = |a_4|^2, \quad |a_2|^2 = |a_5|^2
 \ee
which is easily satisfied.
By computing the gauge generators that are preserved, one can check
that this bound state indeed inherits an $Sp$-projection.

 For $F_b  + F_{b'}$ we consider the
following VEVs:
\be X_{13} = X_{14} = \left(
  \begin{array}{cccccc}
    e_1 & 0 & 0 & 0 & 0 & 0\\
    0 & e_1 & 0 & 0 & 0 & 0\\
    0 & 0 & 0 & e_1 & 0 & 0 \\
    0 & 0 & 0 & 0 & e_1 & 0 \\
  \end{array}
\right), \quad X_{23} = X_{24} = \left(
  \begin{array}{cccccc}
    0 & 0 & e_1 & 0 & 0 & 0 \\
    0 & 0 & 0 & 0  & 0 & e_1 \\
  \end{array}
\right),
\ee
\be
\quad X_{35} = \left(
  \begin{array}{cccccc}
    e_1 & 0 & 0 & 0 & 0 & 0 \\
    0 & e_1 & 0 & 0 & 0 & 0 \\
    0 & 0 & e_1 & 0 & 0 & 0 \\
    0 & 0 & 0 & 0 & 0 & 0 \\
    0 & 0 & 0 & 0 & 0 & 0 \\
    0 & 0 & 0 & 0 & 0 & 0 \\
  \end{array}
\right), \quad X_{45} =
\left(
  \begin{array}{cccccc}
    0 & 0 & 0 & e_1 & 0 & 0\\
    0 & 0 & 0 & 0 & e_1 & 0\\
    0 & 0 & 0 & 0 & 0 & e_1\\
    0 & 0 & 0 & 0 & 0 & 0 \\
    0 & 0 & 0 & 0 & 0 & 0 \\
    0 & 0 & 0 & 0 & 0 & 0 \\
  \end{array}
\right), \ee
\be \quad X_{36} = \left(
  \begin{array}{cccccc}
    0 & 0 & 0 & 0 & 0 & 0 \\
    0 & 0 & 0 & 0 & 0 & 0 \\
    0 & 0 & 0 & 0 & 0 & 0 \\
    0 & 0 & 0 & 0 & 0 & e_1 \\
    0 & 0 & 0 & e_1 & 0 & 0 \\
    0 & 0 & 0 & 0 & e_1 & 0 \\
  \end{array}
\right), \quad X_{46} = \left(
\begin{array}{cccccc}
    0 & 0 & 0 & 0 & 0 & 0\\
    0 & 0 & 0 & 0 & 0 & 0\\
    0 & 0 & 0 & 0 & 0 & 0 \\
    0 & 0 & e_1 & 0 & 0 & 0 \\
    e_1 & 0 & 0 & 0 & 0 & 0 \\
    0 & e_1 & 0 & 0 & 0 & 0 \\
  \end{array}
\right),
\ee
with the remaining VEVs determined by (\ref{altDP5orient}).
Here we took the gamma matrices to be
\be
\gamma_1 =
\left(
  \begin{array}{cc}
     0  & {\bf 1}_{2\times 2} \\
    -{\bf 1}_{2\times 2} & 0 \\
  \end{array}
\right), \qquad
\gamma_2 =
\left(
  \begin{array}{cc}
     0  & 1 \\
    -1 & 0 \\
  \end{array}
\right).
\ee
The
D-terms are satisfied. Note that we have used the notation $F_b
+F_{b'}$ to indicate that this representation has two unbroken
$U(1)$'s. They get mapped into each other under the orientifolding.

After orientifolding the quiver generated by $\{ 2 F_a, 3 F_3,F_4,
F_b + F_{b'}, F_7, 3 F_8 \}$ we get a quiver with the expected
chiral matter content of the MSSM, with three generations, and Higgs
fields. These are some of the most complicated bound states in this
paper, and we have not been able to check that all excess non-chiral
matter can be lifted by an induced superpotential.

We can also argue that all the remaining $U(1)$'s couple to an
independent St\"uckelberg field. This is not automatically true but
can be checked in this case. To see this, before Higgsing and
orientifolding the St\"uckelberg couplings are of the form
\be \sum_{\rm{nodes}\ i} \int C^{(i)} \wedge {\rm Tr}(F_{(i)}), \ee
where $C^{(i)}, F_{(i)}$ are the St\"uckelberg 2-form field and
gauge field for the $i$th node. After Higgsing we get
\be \sum_{i,j}  n_i^j\ \int C^{(i)}\wedge {\rm Tr}({\tilde F}_{(j)})
\ee
where $n_i^j$ is the number of original fractional branes of type
$i$ contained in the bound state $j$, $j \in \{ a,3,4,b,b',7,8 \}$,
and ${\tilde F}_{(j)}$ is the corresponding gauge field. Now it is
easy to check that for our MSSM configuration the rank of $n_i^j$ is
maximal, so that all the $U(1)$'s couple to an independent
St\"uckelberg field. We conclude that the $U(1)_{B-L}$ gauge boson
can be lifted through the St\"uckelberg mechanism.

\newpage

\newsection{Dynamical SUSY breaking}

Since Del Pezzo quivers are chiral, one may expect to find examples
of local models with dynamical supersymmetry breaking. However it
was typically found in examples that if SUSY breaking occurred there
was some runaway mode which invalidated the model
\cite{Berenstein:2005xa,Franco:2005zu,Bertolini:2005di,Intriligator:2005aw,Brini:2006ej}.
Some effort has gone in to finding a way to stabilise such runaway
modes \cite{Franco:2006es,Florea:2006si,Diaconescu:2007ah}.

We have seen that orientifolding eliminates K\"ahler moduli, so one
may revisit this issue by looking for simple models where the
runaway mode is projected out. In fact the new techniques allow us
to engineer many familiar models which are known to break SUSY
dynamically. Examples include:

\newsubsection{A non-calculable model}

\begin{figure}[th]
\begin{center}
            \scalebox{.8}{
               \includegraphics[width=\textwidth]{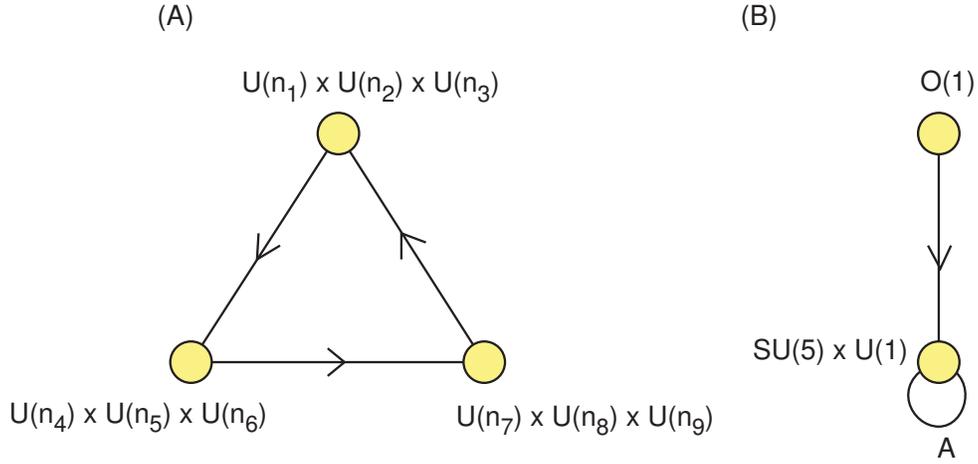}
               }
\end{center}
\vspace{-.5cm} \caption{ \it (A): The quiver for ${\bf C}^3/Z_3
\times Z_3$, an orbifold limit of DP6. (B): Orientifold of a
fractional brane of (A).}\label{ADSnc}
 \end{figure}

Let us consider the ${\bf C}^3/Z_3 \times Z_3$ orbifold. This is the
Calabi-Yau three-fold defined by the equation
\be xyz = t^3, \ee
i.e. it is a cone over a DP6 surface which itself has three $A_2$
singularities. The quiver in shown in figure \ref{ADSnc}A. For
completeness, let us also mention an exceptional collection:
\be
\begin{array}{lll}
  1.\ (1,E_4,-\half) \quad & 4.\ -(2,H,-\half) & 7.\ (1,H\!-\!E_1,0) \\
  2.\ (1,E_5,-\half) & 5.\ -(2,2H\!-\!E_1\!-\!E_2\!-\!E_3,-\half) \quad & 8.\ (1,H\!-\!E_2,0) \\
  3.\ (1,E_6,-\half) & 6.\ -(2,E_4\!+\!E_5\!+\!E_6,-{3\over 2}) & 9.\ (1,H\!-\!E_3,0) \\
\end{array}
\ee
%
%
Now we are interested in the canonical orientifold action on the
nodes, which exchanges oppositely twisted sectors\footnote{This is
the orientifold action we intended to use in version 1 of this
paper, by analogy with the ${\bf C}^3/{Z_3}$ orientifold of section
3.2, but the picture in v1 showed a different orientifold action. It
is not hard to see that with the latter action, with an
antisymmetric projection of the rank 2 tensors and an orthogonal
projection on fixed nodes, the orbifold superpotential would not be
invariant, though we are of course allowed to deform the
superpotential.} :
\be 1 \to 1^*, \quad  2 \to 3^*, \quad 4 \to 9^*,\quad 5 \to
8^*,\quad 6 \to 7^* \ee
%
%
 \begin{figure}[th]
\begin{center}
            \scalebox{.3}{
               \includegraphics[width=\textwidth]{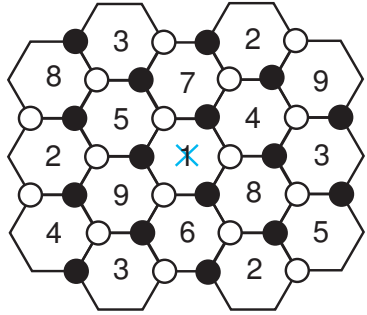}
               }
\end{center}
\vspace{-.5cm} \caption{ \it The dimer for the ${\bf C}^3/Z_3\times
Z_3$ orbifold. The orientifold acts by a $180^0$ rotation centered
at the cross.}\label{DP6Dimer}
 \end{figure}
This is a symmetry of the dimer diagram, as indicated in figure
\ref{DP6Dimer}. Let's take the fractional brane which only uses
nodes $\{ 1,5,8\}$, as shown in figure \ref{ADSnc}. This is a model
with two gauge groups, $U(5)$ and $O(1) = \{\pm 1 \}$. The $U(1)
\subset U(5)$ gets a mass through the St\"uckelberg mechanism, so we
are left only with the $SU(5)$, with matter in the $10 + \bar{5}$.
Since there are no gauge invariant baryonic operators we can write
down, integrating out the massive $U(1)$ leads to a D-term potential
for the dynamical FI-term (a normalizable closed string mode),
stabilizing it. Thus what we are left over with is precisely the
model considered by \cite{Affleck:1983vc}. This model has no
classical flat directions and a non-anomalous R-symmetry that was
argued to be broken, and therefore supersymmetry is broken
dynamically.

An important property of this gauge theory is that it has very few
parameters, so there is little room for a runaway of the parameters
after coupling to 4D gravity. Moreover coming from such a simple
singularity, the model should not be so hard to embed in a compact
CY. For instance, we can easily embed the singularity in the
quintic, by taking an equation of the form
\be 0 = s^2 (xyz-t^3) + x^5 + x^4 s + \ldots \ee
or we could try to use $T^6/Z_3 \times Z_3$. To complete the
analysis we would need to check that the orientifold can be extended
globally and tadpoles can be cancelled.


It would be nice to see the supersymmetry breaking from a dual
gravity perspective. There is presumably an enhan\c{c}on type of
effect at work, similar to \cite{Johnson:1999qt,Wijnholt:2001us}.

\newsubsection{The 3-2 Model}

\begin{figure}[th]
\begin{center}
            \scalebox{.8}{
               \includegraphics[width=\textwidth]{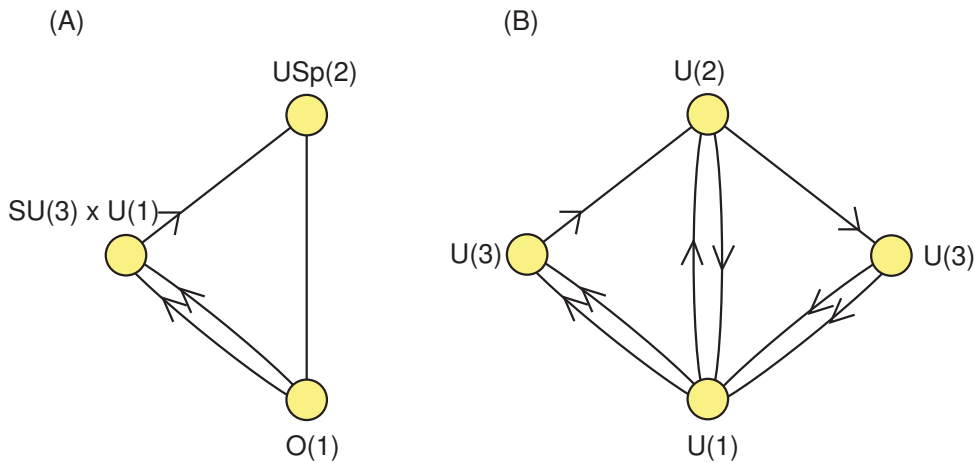}
               }
\end{center}
\vspace{-.5cm} \caption{ \it (A): The 3-2 model, with an extra
massive $U(1)$. (B): Oriented cover of the quiver in
(A).}\label{32Model}
 \end{figure}

Our next model is a little harder to produce, so we start by drawing
the quiver diagram in figure \ref{32Model}A, and its oriented cover
in figure \ref{32Model}B.
%
This is known as the 3-2 model \cite{Affleck:1984uz}. The stringy
version has an additional anomalous $U(1)$, which does not affect
the low energy dynamics as in the previous example. There are
various large $N$ generalizations which appear to have no classical
flat directions and a spontaneously broken R-symmetry, and so should
also break supersymmetry.

The covering quiver has a certain similarity with DP5. So
we take the DP5 quiver with the projections (\ref{DP5proj}) and $a = b = -1$,
with $Sp$-projections on nodes 1 and 5, and $SO$-projections on nodes 2 and 6.
We take
\be \gamma_2 = {\bf 1}_{3\times 3}, \quad \gamma_6 = {\bf
1}_{5\times 5}\ee
and consider a bound
state $F_a$ with charge vector
\be {\rm ch}(F_a) = \sum_i n_i\, \ch{F_i}, \qquad
\vec{n} = \{ 0,3,2,3,0,5,3,2\}.
\ee
Concretely, the VEVs are given by
\be
X_{23} =
\left(
  \begin{array}{cc}
    a_1 & 0 \\
    0 & a_2 \\
    0 & 0  \\
  \end{array}
\right)
, \quad
X_{24} =
\left(
  \begin{array}{ccc}
    a_1 & 0 & 0 \\
    0 & a_2 & 0  \\
    0  & 0  & a_2 \\
  \end{array}
\right) , \ee
\be
 X_{36} = \left(
  \begin{array}{ccccc}
    a_3 & 0 & a_4 & 0 & a_5 \\
    0 & a_6 & 0 & a_7 & 0 \\
  \end{array}
\right)
, \quad
X_{46} =
\left(
  \begin{array}{ccccc}
    0 & a_8 & 0 & 0 & 0  \\
    0 & 0 & a_9 & 0 & 0 \\
    0 & 0 & 0   & 0 & a_{10} \\
  \end{array}
\right),
\ee
with the remaining VEVs either zero or determined by
(\ref{DP5proj}). The D-terms then imply
\be |a_1|^2 = |a_3|^2 + |a_4|^2 + |a_5|^2 = |a_8|^2, \quad |a_2|^2 =
|a_6|^2 + |a_7|^2 = |a_9|^2 = |a_{10}|^2 \ee
which is easily satisfied.

The quiver generated by $\{ 2\, F_1, 3\, F_3, F_a, 3\, F_8 \}$ is of
the required form in figure \ref{32Model}, up to non-chiral matter.
The lepton doublets can get masses only in pairs. Our orientifolded
quiver has five lepton doublets, and in principle we may turn on
mass terms for four of them, leaving one massless.

\newsubsection{ISS meta-stable models}

A number of realizations of ISS vacua \cite{Intriligator:2006dd}
from quivers have already been considered
\cite{Ooguri:2006pj,Argurio:2006ny}. We would like to suggest an
alternative realization, in which the quark masses are obtained in a
more straightforward way.

As we discussed in section {\it 2} we can take the conifold quiver
and turn on mass terms:
\be W = m_{ij}\, {\rm Tr}(A^i B^j) + \lambda\, {\rm Tr}(A^1 B^1 A^2
B^2 - A^1 B^2 A^2 B^1) \ee
We introduce the dimensionless parameters $a = m/E, b = \lambda E$
and consider the regime $a > > b$ so that we can ignore the quartic
term. The non-trivial $U(1)$ gauge group is taken to decouple from
the low energy physics, either by working on the infinite cone or by
coupling to a St\"uckelberg field in compactified settings. We also
set the gauge coupling of $SU(n_2)$ to be very weak. Finally we take
$n_1/2 < n_2 < 3n_1/4$. Then the $SU(n_1)$ flows to strong coupling
and we apply a Seiberg duality. The dual theory has gauge group
$SU(2 n_2-n_1) \times SU(n_2)$, with two pairs of bifundamentals of
opposite charges, and four adjoints for $SU(n_2)$. In particular
both gauge groups are now IR free.

Thus now we are in the situation of ISS, except that we have gauged
a slightly different subgroup of the global symmetry group when the
gauge coupling for $SU(n_2)$ is finite. In this theory SUSY is
broken by the rank condition, and there are meta-stable vacua for
zero adjoint VEVs with pseudomoduli lifted by a one-loop potential.
Actually if we would have kept the quartic terms of the conifold
quiver then we get mass terms for the mesons and we can solve the
F-terms, but because we took $a > > b$ these SUSY vacua are very far
out in meson field space and don't affect the analysis near the
origin.

In the meta-stable vacua, the gauge group $SU(2 n_2-n_1) \times
SU(n_2)$ is broken to $SU(2 n_2-n_1) \times SU(n_1-n_2)$. If the
$SU(n_2)$ gauge coupling is small enough, then both the gauge
couplings of the remaining $SU(2n_2-n_1) \times SU(n_1-n_2)$ are
also very small. There are also some massless Goldstone bosons left
from the broken global symmetries, and some light fermions. The
Goldstone bosons parametrize a compact coset $G/H$ and are not
charged under the remaining gauge groups. Thus the remaining gauge
groups may eventually become strong in the deep IR and generate some
vacuum energy. But since the strong coupling scale is arbitrarily
small, this means there must still be long-lived meta-stable vacua
close to those found by ISS.

\newpage

\newsection{Topological criteria for unification}

From the bottom-up perspective, there is no a priori relation
between the MSSM couplings. For instance the differences between the
(inverse squared) gauge couplings depend on complexified K\"ahler
moduli. If such moduli extend to the UV completion, we will have to
find a suitable potential to stabilize them, and it seems there is
no natural reason to expect any relation between them.

On the other hand we clearly do not live at a random point on the
parameter space. There are many relations among the couplings that
we believe to be a reflection of new physics. So one may ask if
these relations have a special significance in our set-up.

Recently it has been argued that moduli corresponding to
non-normalizable closed string modes may be trivializable, in the
sense that they appear to exist locally but may not be extended to
the UV completion \cite{Buican:2006sn}. (A similar scheme for the
6-volume was proposed in \cite{Becker:2006ks}). This is really a
rephrasing of the obvious fact that the most interesting UV
completions are those which are as rigid as possible, consistent
with observed low energy physics, because they give a topological
explanation of the tree level relations between certain couplings,
as opposed to a dynamical one due to moduli stabilisation.
It also reduces the number of global tadpoles to be cancelled.

Let us review some aspects of this trivialization for K\"ahler
moduli. Suppose that two fractional branes wrap vanishing cycles $a$
and $b$, giving rise to a gauge group $U(n_a) \times U(n_b)$,
embedded in some orientifold compactification. Locally the homology
class $a-b$ has an even lift and an odd lift, where even and odd
refer to the eigenvalue of the homology classes under the
orientifold action. Let us further assume that $a-b$ is the class of
a 2-cycle that does not intersect a vanishing 4-cycle. Then if the
odd lift is trivializable, we have the tree level relation
\be  {4\pi\over g_a^2 }- i {\vartheta_a\over 2\pi} = {4\pi\over
g_b^2} - i {\vartheta_b\over 2\pi }
\ee
and if the even lift is trivializable we have
\be \zeta_a + i c_a = \zeta_b + i c_b \ee
where $\zeta$ couples as an FI-term and $c$ is its axion partner. A
certain linear relation between the axions is needed to keep
hypercharge massless.

Now suppose for ease of discussion that both lifts are
trivializable, so that $a$ and $b$ are the same cycle homologically.
We can model this by considering a single fractional brane with an
adjoint scalar field whose superpotential has two critical points.
Thus morally the vacuum where the two branes sit apart is a Higgsed
vacuum of a unified theory with gauge group $U(n_a + n_b)$, in
particular the tree level gauge couplings of $U(n_a)$ and $U(n_b)$
are related as above. Integrating out the massive adjoint we
generate certain higher order superpotential couplings suppressed by
the mass of the adjoint field, which should correspond to the
subleading complex structure deformations discussed in
\cite{Buican:2006sn}.

This also explains the origin of the monopoles discovered in
\cite{Verlinde:2006bc}: they are D-branes stretched between the
gauge branes, i.e. they are the monopoles in the Higgsed phase
expected by a Pati-Salam like unification in the model of
\cite{Verlinde:2005jr}. The unified model is actually one of the
standard quivers for ${\bf P}^2$, with gauge group $U(4)\times
U(2)\times U(2)$ and $(3,6,3)$ arrows between them.

Let us discuss how these ideas can be applied to the MSSM models
that we have constructed from Del Pezzo 5. For the quivers and
labelling of the nodes we refer to the figures in section {\it 1.2}.
The moduli controlling the difference between the gauge couplings of
$U(3)$ node and the $U(1)_L$ node are trivialisable, so we can give
a topological explanation of the relation $g_3^{-2} = g_1^{-2}$. We
can achieve this by making sure that $E_4-E_5$ or $H - E_1 - E_2 -
E_3$ is globally trivial. If we assume that both are globally
trivial, then we can morally think of the $U(3)$ and $U(1)$ as being
unified in the Pati-Salam group $U(4)$ (i.e. we have baryon-lepton
unification). In fact with only a small change in interpretation
this situation was already proposed in
\cite{Blumenhagen:2003jy,Antoniadis:2000en}.

The further relations\footnote{Due to different normalizations
of abelian and non-abelian charges, for model II equal tension of the branes
corresponds to  $g_2 = \sqrt{2} g_0 = g_3$.}
$g_2 = g_0 = g_3$ look quite natural because they corresponds to the branes
having equal tension, and
they also give precisely the standard tree level relations from $SU(5)$ unification for the
strong, weak and hypercharge couplings \cite{Blumenhagen:2003jy}. But they cannot be imposed by topological
means, because the intersection numbers of the corresponding cycles
are different. However we can do the following. The tree level gauge
coupling $g_3$ corresponds to
\be
{2\over g_3^2} = {N\over g_s \ell_s^2} \int_\alpha B_2, \qquad \alpha = E_4-E_3
\ee
which is a non-normalizable mode in the local geometry. Here $N$ is a numerical
factor which depends on the periodicity of $B_2$. Similarly\footnote{For
model II this would read $\half g_0^{-2} + g_2^{-2} = N g_s^{-1}\ell_s^{-2}\int_\beta B_2$.}
\be\label{trivialized}
{1\over g_2^2} + {1\over g_0^2} = {N\over g_s \ell_s^2}\int_\beta B_2
\ee
is a non-normalizable mode, where $\beta$ is some degree zero linear
combination of the 2-cycles which depends on how we exactly
constructed the bound states\footnote{Actually this relation is not
quite true with the bound states we constructed earlier. However it
would have been true had we avoided the fractional brane $F_6$ in
our bound states, at the cost of making the bound states slightly
more complicated, and changing the orientifold projection in the
case of model II. Alternatively we could replace nodes $1$ and $3$
by bound states which include $F_6$.}. So any linear combination of
these quantities is the integral of the $B$-field over some degree
zero homology class, and therefore potentially trivializable. Together
with $g_3 = g_1$ this then leads to a tree level relation between
the observed gauge couplings
\be\label{toprelation}
{1\over g_Y^2} + {1\over g_W^2} =
\left[{2 n_1 \over n_2} + {2\over 3}\right] {1\over
g_S^2} \ee
where we have assumed that $n_1 \alpha - n_2 \beta $ is trivialized.
This is compatible with the
relations from $SU(5)$ unification, $g_S^{-2} = g_W^{-2} = {3\over
5} g_Y^{-2}$ if we take $\beta =
\alpha$ in homology.

Since $\int_\alpha B_2 - \int_\beta B_2 = \int_\gamma H_3$, where
$\del \gamma =\alpha - \beta$, one might wonder if these tree level
relations are not affected if we turn on background
fluxes\footnote{We would like to thank Angel Uranga for pointing out
this possibility.}. This seems unlikely for the following reason. As
long as no background fields are turned on we should expand the
$B$-field in harmonic forms and the above criterion is sufficient to
guarantee that the gauge couplings are related. When we turn on
general closed string deformations it is not necessarily true that
the $B$-field must be expanded in harmonic forms; however it is
unlikely that we gain additional zero modes and so the relation
between the couplings, which is due to a lack of zero modes, should
not be affected.


\newpage

\newsection{Final thoughts}\label{finalsection}

It should be clear by now that the set of quiver gauge theories that
can be obtained from branes at singularities is rather large. It is
our impression that virtually any quiver can be constructed locally,
and we believe this is really the main message of this paper. It
will probably not be possible to couple every such model to 4D
gravity, but given a local model it will be very hard to argue that
it {\it cannot} be consistently coupled to gravity.

One would like to understand if the local D-brane scenario has
something to add to discussions of beyond the SM physics. On the one
hand it seems premature. After many years of work, the field theory
community has not been able to come up with a single model that
adresses all concerns. String phenomenology is going to have to
address the same issues, and barring a miraculous discovery of the
correct UV completion it would be naive to expect that doing string
phenomenology would magically improve on this situation. On the
other hand, string phenomenology has traditionally been more a
source of new ideas and intuition than a source of accurate models.

 One way to
proceed is to try and isolate desirable features and translate them
into geometrical or even topological terms. We have already
discussed a topological explanation for tree level relations between
gauge couplings. Other important issues are flavour problems. For
instance we would like to explain hierarchies among the Yukawa
couplings, and we would like to explain why new physics doesn't
generate large FCNC's. Can we translate these criteria into
geometric terms, and perhaps guarantee them through topological
mechanisms similar to section {\it 6}? Such rigidity requirements
may eliminate the majority of UV completions in the string
landscape.

Another important criterion will be stability. The apparent
long-lived nature of our universe suggests we are in a vacuum which
does not have too many vacua in its neighbourhood with a large
cumulative probability to decay away to.\footnote{In fact it has
recently been argued \cite{Nima} in a `bottom-up' approach, quite
independent of string theory, that the landscape of the SM plus
quantum gravity may contain vacua close to ours which correspond to
compactification to lower dimensions.} This makes stability a very
acute issue with possible predictive power, and we might expect
string theory to have something to say about this.

The bottom-up perspective also allows us to take a step back and see
if top-down approaches could benefit from new ingredients. It seems
that non-commutative internal geometries play an important role.
Thus perhaps we should be paying more attention to backgrounds of
the type recently constructed in \cite{Grana:2006kf}.

\newsubsection{On soft SUSY breaking}

In this paper we have repeatedly used the technique of deforming the
gauge theory to get rid off undesirable particles. By the same logic
we can proceed to turn on masses for all the superpartners of the
Standard Model fields and recover the non-SUSY Standard Model
itself. This must be possible, but it is not terribly helpful.
Although the dictionary between the local geometry and the
superpotential deformations used in this paper are fairly well
understood and can in principle be solved exactly, unfortunately the
dictionary for SUSY breaking deformations is more complicated. It
requires understanding Kaehler deformations which can probably only
be addressed numerically, and it requires further generalizing the
notion of geometry in ways that are probably not yet quite
understood. Even for simpler theories like $N=4$ Yang-Mills, little
is known about the stringy description of non-supersymmetric
deformations. Progress can perhaps be made if one can identify
special points on the parameter space where the stringy description
simplifies.

%

\newsubsection{Composite Higgses?}

A rather curious feature of the MSSM quiver is that, if we started
without Higgs fields, we can automatically generate them through
Seiberg duality on $U(3)_c$ or $U(1)_L$. For the $U(3)$ node,
Seiberg duality has been considered by Matt Strassler
\cite{Strassler}, who was looking for a possible embedding into a
duality cascade, and also in \cite{Maekawa:1995cz}. A problem in
that case is that one needs to add extra massive matter in order to
make the $SU(3)$ coupling grow strong towards the UV.

In the MSSM quivers we have an alternative: we may consider a
``Seiberg duality'' on the $U(1)_L$ node (node $1$ in the figures of
section {\it 1.2}). The dual group is $U(5)$ or larger and we are
automatically have $N_f = N_c + 1$, so we don't have to add
additional matter to get a consistent picture. Since the $U(1)
\subset U(5)$ and the $SU(5) \subset U(5)$ run independently, this
may even be consistent with unification, but we haven't done the
calculation. As the $SU(5)$ flows to strong coupling towards the IR,
the electric `quarks' bind into mesons which have the same quantum
numbers as the Higgses, although the large number we get is not so
desirable. Their number may be reduced if we also have Higgs fields
with appropriate couplings in the electric theory. Thus perhaps if
the supersymmetry breaking scale is significantly lower than the
scale at which we would have to apply a Seiberg duality, the Higgses
may be interpreted as composite fields, and this might be the seed
of an explanation why one pair ends up being relatively light.

\newsubsection{The QCD string as a fundamental string}

One may wonder if our set-up gives any insight into the stringy
description of QCD. Let us turn on Higgs VEVs so that the quarks
obtain small masses. Then in the IR we may focus on the $U(3)$ node
which gives us pure $SU(3)$ SUSY QCD. Now note that the $U(3)$ brane
(together with its orientifold image) has wrapping numbers
$(0,E_4-E_3,0)$. So as the theory flows to strong coupling what most
likely happens is that the Del Pezzo undergoes a conifold
transition, where a small 2-sphere in the class $E_4 - E_3$ gets
replaced by a 3-sphere. After the transition, the $U(3)$ brane has
been replaced by flux, and thus the open strings ending on this
brane are confined. The glueball condensate is described by a closed
string mode, as envisaged in \cite{Klebanov:2000hb}.

Thus in our picture the graviton and the QCD string can be described
as different modes of the same fundamental string, the IIB string.
As we discuss momentarily though, in some situations the graviton is
better described as a mode of the heterotic string.

\newsubsection{Weakly coupled Planck brane?}

As we reach the Planck brane, we will start to see other ingredients
of the compactification: D7's and orientifold planes (which split up
non-perturbatively as mutually non-local D7-branes). Generically the
IIb string coupling cannot be kept small, and the IIB description
may be less than useful. However there may exist a large class of UV
completions where the Planck brane has a dual description in terms
of weakly coupled heterotic strings.

The fractional brane configuration only has net D3 and D5 charge, so
it should get mapped to a small (possibly constrained) instanton in
the heterotic string. As the instanton shrinks to zero, it generates
a throat and thus potentially a large hierarchy. This is the
heterotic manifestation of the decoupling limit. The heterotic
dilaton grows down the throat, so the MSSM is non-perturbative from
this point of view.

In the perturbative heterotic description we can not see the
enhanced gauge symmetry due to fractionation, since the individual
fractional branes have D7-brane charges. Thus the fractional branes
should have merged into a single NS5-brane when the heterotic
coupling is small. This by itself does not mean that the gauge
couplings should unify at the cross-over scale, although it seems
like a natural boundary condition.

\bigskip
\noindent {\it Acknowledgements}:

It is a pleasure to thank Sebastian Franco, Sam Pinansky and Herman
Verlinde for discussions. I would also like to thank CERN and the
Max Planck Institute in Munich for hospitality and the opportunity
to present some of these results.

\end{document}